\documentclass[review]{elsarticle}

\usepackage{amssymb, amsmath, amsfonts, setspace, epsfig, graphicx, psfrag, booktabs, subfig}

\biboptions{semicolon}

\usepackage[normalem]{ulem}

\usepackage{color}

\def\beq{\begin{equation}}
\def\eeq{\end{equation}}

\newcommand{\bma}[1]{\mbox{\boldmath $#1$}}

\def\XXint#1#2#3{{\setbox0=\hbox{$#1{#2#3}{\int}$}
     \vcenter{\hbox{$#2#3$}}\kern-.52\wd0}}       

\journal{...}

\begin{document}

\begin{frontmatter}

\title{A boundary perturbation method to simulate nonlinear deformations of a two-dimensional bubble}

\author{Philippe Guyenne}

\address
{Department of Mathematical Sciences, University of Delaware, Newark, DE 19716, USA}
\ead{guyenne@udel.edu}




\begin{abstract}
Nonlinear deformations of a two-dimensional gas bubble are investigated
in the framework of a Hamiltonian formulation involving surface variables alone.
The Dirichlet--Neumann operator is introduced to accomplish this dimensional reduction 
and is expressed via a Taylor series expansion.
A recursion formula is derived to determine explicitly each term in this Taylor series up to an arbitrary order of nonlinearity.
Both analytical and numerical strategies are proposed to deal with this nonlinear free-boundary problem
under forced or freely oscillating conditions.
Simplified models are established in various approximate regimes,
including a Rayleigh--Plesset equation for the time evolution of a purely circular pulsating bubble,
and a second-order Stokes wave solution for weakly nonlinear shape oscillations that rotate steadily on the bubble surface.
In addition, a numerical scheme is developed to simulate the full governing equations,
by exploiting the efficient and accurate treatment of the Dirichlet--Neumann operator via the fast Fourier transform. 
Extensive tests are conducted to assess the numerical convergence of this operator as a function of various parameters.
The performance of this direct solver is illustrated by applying it to the simulation of cycles of compression-dilatation
for a purely circular bubble under uniform forcing,
and to the computation of freely evolving shape distortions represented by steadily rotating waves and time-periodic standing waves.
The former solutions are validated against predictions by the Rayleigh--Plesset model,
while the latter solutions are compared to laboratory measurements in the case of mode-2 standing waves.
\end{abstract}


\end{frontmatter}


\section{Introduction}

Gas bubble dynamics has been the subject of intensive research in recent decades due to its prominent role
in a number of physical phenomena.
For example, bubble formation during gas entrainment into water by breaking waves produces sound over large distances
which contributes to ambient noise in the ocean.
Oscillations of bubble volume may lead to collapse and generate shock waves with damaging effects on the submerged machinery
in cavitation problems.
Large distortions of bubble shape may lead to break-up and contribute to increasing bubble population in the surrounding fluid.
Related phenomena such as sonoluminescence \cite{bhl02}, and applications involving the ultrasonic excitation of microbubbles
for medical imaging \cite{l04} or the implosion of deuterium-tritium capsules for inertial confinement fusion \cite{hurricane14},
have drawn a lot of attention in recent years.
We are aware of a similar large literature on the dynamics of fluid droplets, which is closely related to the present subject,
but for convenience we will restrict the following introduction to gas bubbles.

In theoretical studies of cavitation for a single bubble, a common approach relies on the Rayleigh--Plesset (RP) model 
which is a second-order nonlinear ordinary differential equation (ODE) describing the time evolution of the bubble radius
under the influence of surface tension, viscosity and an external pressure field.
In general, the latter is meant to represent a source of disturbances from the surrounding environment
(e.g. internal variations of the fluid flow, perturbations due to atmospheric pressure or waves at the sea surface, 
noise generated by submerged machinery or by other bubbles).
The resulting oscillations of the bubble radius are usually referred to as radial or volume oscillations.
The RP model can be derived from the incompressible Navier--Stokes equations 
by assuming spherical symmetry (i.e. angular invariance) with appropriate boundary conditions.
Its simple closed form as compared to the full Navier--Stokes equations makes it well suited for
both mathematical analysis and numerical simulation.
In particular, bifurcation theory for nonlinear ODEs has been successfully applied to characterizing 
the rich behavior of solutions under external forcing.
Reviews on this model with a detailed discussion on its properties, predictions and extensions
(incorporating e.g. compressible or thermal effects) can be found in \cite{fl97,pp77,p17}.

For radial pulsations of sufficiently large amplitude, distortions of the bubble shape 
(with angular variation) tend to develop, induced by parametric instability as reported in experimental observations.
Investigation of these shape oscillations has also produced an abundant literature,
with earlier mathematical results focusing on the linear case for small distortions \cite{pp77}
or the weakly nonlinear case without radial motion \cite{tb83}.
The problem of bubble dynamics with a nontrivial geometry is significantly more difficult to analyze than the RP framework,
and the vast majority of theoretical studies have considered three-dimensional axisymmetric bubbles in potential flow,
meaning that their shape is assumed to be invariant with respect to the azimuthal angle in spherical coordinates.
Seeking a weakly nonlinear solution in the form of an asymptotic series for small- to moderate-amplitude distortions
and employing the method of multiple scales, a hierarchy of inhomogeneous linear equations can be established 
to determine each order of approximation as a function of previous orders.
Then, by expanding each order in terms of Legendre polynomials relative to the inclination angle (owing to the axisymmetry)
and by exploiting their orthogonality property, explicit expressions can be obtained for the shape modes
(i.e. the coefficients in factor of the Legendre polynomials) at any order in the time-periodic setting.
These perturbation calculations are particularly tedious because the inhomogeneous equations become increasingly 
complicated with the level of approximation, and for this reason, they have usually been restricted to a few leading orders.

In earlier work, Tsamopoulos and Brown \cite{tb83} derived such an asymptotic solution up to second order 
in wave amplitude for the first four shape modes in the freely oscillating regime (without radial motion or external forcing).
Their analysis reveals the natural wave frequencies at first order (i.e. in the linear approximation)
as well as effects from nonlinear wave interactions at higher order.
More intriguing phenomena can occur if both volume and shape modes are coupled together,
especially under resonant or near-resonant conditions, with one or more shape modes being excited.
On one hand, radial oscillations driven by an external acoustic field can trigger shape distortions via parametric instability.
This stability problem for a pulsating spherical bubble has been extensively investigated at lowest order in the asymptotic procedure,
where the volume mode typically obeys the RP equation while the distortion modes satisfy a Mathieu-type equation
with non-autonomous coefficients depending on the radial motion \cite{bhl02,pp77}.
Exponential growth is predicted for this instability at the linear level,
but the inclusion of higher-order nonlinear corrections has been found to promote its saturation \cite{gi18,hkn13}.
On the other hand, second-order interactions among shape modes can transfer energy to the volume mode, 
potentially leading to a monopole emission of sound \cite{lh89} or an erratic motion of the bubble \cite{be90,d04}.

Aside from theoretical results based on asymptotic solutions or reduced models, there is also a large literature
on direct numerical simulations of bubble dynamics,
owing to advances in computer power and numerical methods over the last few decades.
Typically solving the Navier--Stokes equations and using interface reconstruction techniques on Cartesian grids, 
these studies have considered a variety of configurations involving multiple bubbles 
and complex flow conditions or complex surrounding domains, e.g. \cite{brgaba23,et99,fp18,gmm22,hl07b} to cite a few references.
While such computations have produced impressive results, being able to handle extreme situations  
such as bubble merging or break-up, they may not be the best option when it comes to accurately simulating 
shape oscillations of a bubble because the computed solution is particularly prone to numerical diffusion.
Moreover, their computational cost is usually high because they require solving the governing equations over the entire physical domain.
If advective inertial effects are assumed to be small compared to viscous effects,
then the Navier--Stokes equations can be simplified to those for Stokes flow
and boundary integral methods have been developed for their simulation,
e.g. in view of medical applications to microbubbles for drug delivery in blood flow \cite{gg18}.
However, the flow linearity associated with this approach makes it inadequate for the description of nonlinear bubble deformations.

Closer to the present problem, direct numerical simulations based on nonlinear potential flow theory
have also been performed via boundary integral and conformal mapping methods \cite{lsvl21,ml99b,p04,tong97,wb10,zl15}.
These exploit coordinate transformations that are motivated by complex analysis or by the bubble geometry,
and they only involve dynamical variables representing the bubble surface.
Both methods can resolve strong shape distortions (without change in topology),
with boundary integral techniques being more adaptable to various flow configurations,
while the conformal mapping approach is restricted to the two-dimensional setting.
In particular, the latter method enables the exact derivation of steady solutions 
or the time-dependent computation of rotating and standing waves 
for nonlinear oscillations on a bubble surface, similar to Crapper's solutions for the capillary water wave problem \cite{c99,d21,wc00}.
Using a series expansion in Legendre polynomials without assuming small amplitudes, 
McDougald and Leal \cite{ml99a} solved numerically the nonlinear system
of ordinary differential equations for the radial and shape modes of a three-dimensional axisymmetric bubble.
However, the computational cost was high and these authors only examined the interaction of a few leading modes.

In this paper, we present a combination of mathematical and numerical results on the surface dynamics of a gas bubble
immersed in a two-dimensional liquid of infinite extent.
Volume and shape oscillations of the bubble are considered under forced or unforced conditions by a far-field pressure.
The starting point is the Hamiltonian formulation for nonlinear potential flow around a deformable bubble 
under the influence of surface tension without gravity, as proposed by Benjamin \cite{b87}.
The corresponding conjugate variables turn out to be natural choices of surface quantities for this free-boundary problem,
thus allowing for dimensionality reduction.
In this theoretical framework, our new contributions include:
\begin{enumerate}
\item Detailed restatement of this Hamiltonian formulation by introducing the Dirichlet--Neumann operator (DNO)
to express the full governing equations as a closed system in terms of surface variables alone.
In doing so, the connection between these governing equations and the conserved Hamiltonian is also clarified.
Specifics of this two-dimensional problem were not treated by Benjamin \cite{b87}.

\item Taylor series representation of the DNO associated with the Laplace problem in polar coordinates.
Each term in this Taylor series is determined explicitly by a recursion formula involving the bubble deformations
relative to a quiescent circular state.

\item Analytical calculation of a second-order Stokes wave solution for steadily rotating shape oscillations 
in the weakly nonlinear regime, without volume change.
Asymptotic expressions are obtained for the surface displacement, velocity potential and angular wave speed.

\item Derivation of a two-dimensional inviscid version of the RP equation in the purely circular geometry.
Its independent Hamiltonian structure and linear dispersion relation for radial pulsations driven by a uniform pressure field
are also established.

\item Development of an accurate and efficient numerical scheme for the direct simulation of nonlinear bubble deformations,
coupling shape oscillations and radial pulsations together.
It is applicable to surface distortions of moderate steepness,
nonetheless the computations can be performed up to an arbitrary order of nonlinearity with an arbitrary number of shape modes
in an automatic manner via the fast Fourier transform.

\item Extensive tests to examine the numerical convergence of the DNO.
Validation against predictions by the RP equation is also provided in the purely circular case under uniform pressure forcing.

\item Computations of nonlinear shape oscillations represented by steadily rotating waves and time-periodic standing waves 
on the bubble surface in the unforced regime without volume change.
For the latter solutions, comparison with existing laboratory measurements is presented 
on the frequency of mode-2 oscillations as a function of the bubble's maximum aspect ratio.
\end{enumerate}
While the DNO has been a common tool in the formulation, analysis and simulation of the water wave problem \cite{cgs21,cs93,g19,nr01,xg09},
its applications have focused mostly on cases where the reference geometry is rectangular with Cartesian coordinates.
We are only aware of a few exceptions. 
De la Llave and Panayotaros \cite{dp96} considered nonlinear gravity waves on the surface of a sphere 
and derived a series expansion for this operator in terms of spherical coordinates.
Their study was strictly mathematical and did not produce any numerical result.
Guyenne and P\u ar\u au \cite{gp16} adopted a similar approach in the axisymmetric cylindrical setting
to compute solitary waves on the surface of a ferrofluid jet.
This formalism has also been applied to scattering problems in acoustics and electromagnetics 
with non-Cartesian (e.g. polar or spherical) parameterizations of the irregular domain \cite{fns07,nn04}.
However, these investigations have been restricted to linear time-harmonic waves in the presence of a stationary object with a fixed shape.
To our knowledge, it is the first time here that the DNO is introduced to describe this nonlinear hydrodynamic problem
with a moving boundary in polar coordinates.
Furthermore, unlike all previous applications, the DNO in this case contains an additional component
to allow for radial pulsations of the bubble (i.e. volume changes such as compression or dilatation).

While the problem under consideration is two-dimensional in space,
this is not a limitation of the mathematical formulation and numerical procedure that we advocate to analyze it.
The present study contributes to their development in this idealized situation
before tackling the more general three-dimensional case in the future.
This two-dimensional problem is challenging and interesting in its own right,
but has not been examined much in previous modeling work
as opposed to e.g. the three-dimensional axisymmetric configuration.
Indeed, experimental observations have revealed possible significant differences in bubble dynamics 
between the two- and three-dimensional geometries,
e.g. regarding the condensation, collapse and sonoluminescence of explosive bubbles \cite{d19}.
Even for the three-dimensional axisymmetric problem which essentially reduces to a planar analysis,
there are qualitative and quantitative differences as compared to the purely two-dimensional case.
The present results may thus serve as benchmark solutions to test other mathematical or numerical models in this area.
Lastly, understanding the dynamics of two-dimensional bubbles may be relevant in applications to microfluidics
where the technology is based on small thin devices (e.g. PDMS chips).
Recent research has investigated ways to remove unwanted air bubbles from microfluidic systems \cite{hwmzw21}
or, on the contrary, to exploit their presence for various purposes such as 
micropumps, micromixers, microvalves and microactuators \cite{kaaysk15}.

The remainder of this paper is organized as follows.
In Section 2, we present the mathematical formulation of this two-dimensional problem on the nonlinear deformations of a gas bubble,
and we elaborate on its Hamiltonian structure by introducing the DNO to accomplish the reduction to surface variables.
In Section 3, we derive a Taylor series representation for the DNO with two distinct components 
to enable both shape distortions and volume variations of the bubble.
In Section 4, we discuss various analytical approximations including a second-order Stokes wave solution
for steadily rotating shape oscillations and a RP model for purely circular pulsations.
In Section 5, we describe the numerical methods for space discretization and time integration
to solve the full governing equations in Hamiltonian form.
In Section 6, we show numerical tests on the convergence of the DNO 
as well as direct computations of rotating and standing waves on the bubble surface.
For standing waves, a comparison with laboratory measurements and other theoretical results is provided.
Compression-dilatation cycles under the excitation of a far-field pressure 
are also simulated and validated against predictions by the RP model.
Finally, concluding remarks are given in Section 7.

\section{Mathematical formulation}

\subsection{Governing equations}
\label{BasicEqs}

We consider the motion of a single gas bubble immersed in a two-dimensional liquid (e.g. water) 
spanning a domain ${\cal D}$ of infinite extent.
Given the geometry of this problem, we adopt a polar coordinate system $(r,\theta)$ 
whose origin coincides with the bubble center such that
\[
{\cal D} = \mathbb{R}^2 \setminus \{ 0 \le \theta < 2\pi, 0 \le r < R + \eta(\theta,t) \} \,,
\]
where $R > 0$ is the radius of the unperturbed circular bubble
and $\eta(\theta,t)$ denotes the surface perturbation relative to this simple configuration at any time $t$.
We disregard any motion of the bubble center and restrict our attention to its surface dynamics,
which implies in particular that body forces such as gravity or buoyancy are neglected.
The exterior flow is assumed to be incompressible, inviscid and irrotational so that the fluid velocity
is given by ${\bf u}(r,\theta,t) = \nabla \varphi$ where the velocity potential $\varphi(r,\theta,t)$ satisfies the Laplace equation
\begin{equation} \label{laplace}
\nabla^2 \varphi = 0 \,, \quad \mbox{in} \quad {\cal D} \,,
\end{equation}
with $\nabla = (\partial_r,r^{-1} \partial_\theta)^\top$ being the spatial gradient in polar coordinates.
At the free surface 
\[
{\cal S} = \{ 0 \le \theta < 2\pi, r = s(\theta,t) = R + \eta(\theta,t) \} \,,
\]
there are two boundary conditions, namely the kinematic condition
\begin{equation} \label{kine}
\partial_t \eta = \partial_r \varphi - s^{-2} (\partial_\theta \eta) (\partial_\theta \varphi) \,,
\quad \mbox{on} \quad {\cal S} \,,
\end{equation}
and the dynamic (or Bernoulli's) condition
\begin{equation} \label{dyna}
\partial_t \varphi = -\frac{1}{2} |\nabla \varphi|^2 + \frac{\sigma}{\rho} \kappa + \frac{1}{\rho} \Delta p \,,
\quad \mbox{on} \quad {\cal S} \,,
\end{equation}
where
\begin{equation} \label{curvature}
\kappa = \frac{s^2 + 2 (\partial_\theta \eta)^2 - s \partial_\theta^2 \eta}{\big( s^2 + (\partial_\theta \eta)^2 \big)^{3/2}} \,,
\end{equation}
denotes the mean curvature at any point on the free surface.
Keep in mind that $\partial_t s = \partial_t \eta$, $\partial_\theta s = \partial_\theta \eta$ 
and similarly for higher derivatives in \eqref{kine}--\eqref{curvature}.
The parameters $\sigma$ and $\rho$ represent the surface tension and fluid density respectively.
Typical values are $\sigma = 75$ dyn cm$^{-1}$ and $\rho = 1$ g cm$^{-3}$ for water.
Because the bubble interior is not empty, the contribution $\Delta p = p_\infty - p_B$ in \eqref{dyna} denotes the difference 
between the fluid pressure $p_\infty$ in the far field and the bubble pressure $p_B$ exerted by the internal gas.
The competition between these two pressure disturbances drives the bubble deformations.
More details on $p_\infty$ and $p_B$ will be provided in a subsequent section.
Finally, the vanishing condition
\begin{equation} \label{farfield}
| \nabla \varphi| \to 0 \,, \quad \mbox{as} \quad r \to +\infty \,,
\end{equation}
is imposed in the far field.
Note that the boundary conditions in $\theta$ are naturally periodic in this geometric configuration.
The present problem is nonlocal and nonlinear due to the dependence on ${\cal S}$ 
and associated boundary conditions \eqref{kine}--\eqref{dyna}.

An interesting situation arises in the absence of pressure disturbances, i.e. $\Delta p = {\rm constant}$, 
which may be set to $\Delta p = 0$ without loss of generality
through a gauge transformation on the velocity potential $\varphi$ \cite{cgs21}.
If so, this system of equations possesses a number of invariants of motion, notably the energy
\begin{equation} \label{ener}
H = \int_{\cal D} \frac{1}{2} |\nabla \varphi|^2 dA + \int_{\cal S} \frac{\sigma}{\rho} d\ell = K + P \,,
\end{equation}
where $dA = r dr d\theta$ is the elementary area over ${\cal D}$ 
and $d\ell = \sqrt{s^2 + (\partial_\theta \eta)^2} d\theta$ is the elementary arclength along ${\cal S}$.
The first term $K$ in \eqref{ener} is the kinetic part associated with fluid motion,
while the second term $P$ is the potential part associated with surface tension.
Other invariants include the mean surface level
\begin{equation} \label{level}
Q = \int_0^{2\pi} s \, d\theta \,,
\end{equation}
the mass (or volume)
\begin{equation} \label{mass}
V = \int_{\cal D} dA \,,
\end{equation}
and the angular momentum (or impulse)
\begin{equation} \label{moment}
I = \int_{\cal D} ({\bf r} \times \nabla \varphi) \, dA \,,
\end{equation}
where ${\bf r} = r \, {\bf e}_r$ denotes the position vector.
It turns out that
\[
I = \int_{\cal D} (\partial_\theta \varphi) \, r dr d\theta = \int_s^{+\infty} \left( \varphi \big|_0^{2\pi} \right) r dr = 0 \,,
\]
by virtue of the periodic boundary conditions in $\theta$.
Furthermore, because the fluid mass as defined in \eqref{mass} is clearly infinite, it may be substituted by its complement 
over the bounded area of the gas bubble, more specifically
\begin{equation} \label{mass2}
V = \int_0^{2\pi} \int_0^s r dr d\theta = \int_0^{2\pi} \frac{1}{2} s^2 d\theta \,,
\end{equation}
which is also conserved over time.

\subsection{Hamiltonian formulation}

As shown by Benjamin \cite{b87}, Eqs. \eqref{laplace}--\eqref{farfield} with $\Delta p = 0$ can be re-expressed as 
a canonical Hamiltonian system
\begin{equation} \label{canonical}
\partial_t \begin{pmatrix}
\eta \\ 
\xi
\end{pmatrix} = \begin{pmatrix}
0 & -1 \\ 
1 & 0
\end{pmatrix} \begin{pmatrix}
\partial_\eta H \\ 
\partial_\xi H
\end{pmatrix} \,,
\end{equation}
in terms of the two conjugate variables $\eta(\theta,t)$ and 
\begin{equation} \label{dirichlet}
\xi(\theta,t) = \varphi(s(\theta,t),\theta,t) \,,
\end{equation}
the latter being the trace of the velocity potential evaluated at the free surface ${\cal S}$.
Note the sign difference in the symplectic matrix of \eqref{canonical} as compared to the standard canonical form
(see e.g. the water wave problem \cite{z68}),
which is explained by the fact that ${\cal S}$ is the inner boundary of the fluid domain ${\cal D}$ with respect to the coordinate system.
The Hamiltonian $H$ in \eqref{canonical} coincides with the energy \eqref{ener}.
Because Benjamin \cite{b87} did not specifically cover this two-dimensional case in polar coordinates,
we will elaborate further on it here.
In doing so, we provide a new perspective on this problem by introducing the Dirichlet--Neumann operator (DNO)
\begin{equation} \label{DNO}
G(\eta) \xi = (-1,s^{-1} \partial_\theta \eta)^\top \cdot \nabla \varphi \big|_{r=s}
= \sqrt{1 + s^{-2} (\partial_\theta \eta)^2} (\nabla \varphi \cdot {\bf n}) \big|_{r=s} \,,
\end{equation}
which is the singular integral operator that takes Dirichlet data $\xi$ on ${\cal S}$, 
solves the Laplace equation \eqref{laplace} subject to \eqref{farfield},
and returns the corresponding Neumann data (i.e. the normal fluid velocity on ${\cal S}$).
The outward unit vector ${\bf n}$ normal to ${\cal S}$ is given by
\begin{equation} \label{normal}
{\bf n} = \frac{(-1,s^{-1} \partial_\theta \eta)^\top}{\sqrt{1 + s^{-2} (\partial_\theta \eta)^2}} 
= \frac{s}{\sqrt{s^2 + (\partial_\theta \eta)^2}} (-1,s^{-1} \partial_\theta \eta)^\top \,,
\end{equation}
which points in the opposite $r$-direction, consistent with the sign difference in the canonical form \eqref{canonical}
as mentioned above.
The choice of definition \eqref{DNO} for the DNO will become more evident in subsequent calculations.
It is a linear operator in $\xi$ but depends nonlinearly on $\eta$.
Similar to studies on water waves \cite{cs93,g17,gn07,xg09}, an advantage of using the DNO is that the dependence on 
$\eta$ and $\xi$, which are the two conjugate variables, is made more explicit
in the equations of motion \eqref{kine}--\eqref{dyna} and in the Hamiltonian \eqref{ener}.
Indeed, the variable $\eta$ appears as part of the domain of integration ${\cal D}$
in the original form \eqref{ener} of $H$, while the variable $\xi$ does not even appear explicitly there,
and so it is not clear at this stage what the variational derivatives of $H$
with respect to $(\eta,\xi)$ mean as indicated by \eqref{canonical}.

In the following, we will write \eqref{canonical} more explicitly via the DNO.
With the definitions \eqref{dirichlet} and \eqref{DNO} at hand, we get the identities
\begin{equation} \label{Phit}
\partial_t \xi = \partial_t \varphi + (\partial_t \eta) (\partial_r \varphi) \big|_{r=s} \,, \quad
\partial_\theta \xi = \partial_\theta \varphi + (\partial_\theta \eta) (\partial_r \varphi) \big|_{r=s} \,, 
\end{equation}
by differentiating \eqref{dirichlet} and applying the chain rule.
Together with 
\begin{equation} \label{DNO2}
G(\eta) \xi = -\partial_r \varphi + s^{-2} (\partial_\theta \eta) (\partial_\theta \varphi) \big|_{r=s} \,,
\end{equation}
from \eqref{DNO}, we deduce
\begin{equation} \label{Phia}
\partial_\theta \varphi \big|_{r=s} = \partial_\theta \xi - (\partial_\theta \eta) (\partial_r \varphi) \big|_{r=s} \,, 
\end{equation}
and
\begin{eqnarray*}
\partial_r \varphi \big|_{r=s} & = & -G(\eta) \xi + s^{-2} (\partial_\theta \eta) (\partial_\theta \varphi) \big|_{r=s}
= -G(\eta) \xi + s^{-2} (\partial_\theta \eta) \big[ \partial_\theta \xi - (\partial_\theta \eta) (\partial_r \varphi) \big] \big|_{r=s} \,, \\
& = & -G(\eta) \xi + s^{-2} (\partial_\theta \eta) (\partial_\theta \xi) - s^{-2} (\partial_\theta \eta)^2 (\partial_r \varphi) \big|_{r=s} \,,
\end{eqnarray*}
which implies
\begin{equation} \label{Phir}
\partial_r \varphi \big|_{r=s} = \frac{1}{1 + s^{-2} (\partial_\theta \eta)^2} 
\big[ s^{-2} (\partial_\theta \eta) (\partial_\theta \xi) - G(\eta) \xi \big] \,.
\end{equation}
Then substituting \eqref{Phir} back into \eqref{Phit} and \eqref{Phia}, we find
\begin{eqnarray} \label{Phia2}
\partial_\theta \varphi \big|_{r=s} & = & \partial_\theta \xi - \frac{\partial_\theta \eta}{1 + s^{-2} (\partial_\theta \eta)^2} 
\big[ s^{-2} (\partial_\theta \eta) (\partial_\theta \xi) - G(\eta) \xi \big] \,, \nonumber \\
& = & \frac{1}{1 + s^{-2} (\partial_\theta \eta)^2} \big[ \partial_\theta \xi + (\partial_\theta \eta) G(\eta) \xi \big] \,,
\end{eqnarray}
and
\begin{eqnarray} \label{Phit2}
\partial_t \varphi \big|_{r=s} & = & \partial_t \xi - (\partial_t \eta) (\partial_r \varphi) \big|_{r=s} \,, \nonumber \\
& = & \partial_t \xi + \frac{G(\eta) \xi}{1 + s^{-2} (\partial_\theta \eta)^2} 
\big[ s^{-2} (\partial_\theta \eta) (\partial_\theta \xi) - G(\eta) \xi \big] \,,
\end{eqnarray}
where we have used the relation
\[
\partial_t \eta = \partial_r \varphi - s^{-2} (\partial_\theta \eta) (\partial_\theta \varphi) \big|_{r=s} = -G(\eta) \xi \,,
\]
according to the kinematic condition \eqref{kine} and the definition \eqref{DNO2} of the DNO.
Combining the squares of \eqref{Phir} and \eqref{Phia2}, i.e.
\begin{eqnarray*}
(\partial_r \varphi)^2 \big|_{r=s} & = & \frac{1}{\big( 1 + s^{-2} (\partial_\theta \eta)^2 \big)^2} 
\Big[ s^{-4} (\partial_\theta \eta)^2 (\partial_\theta \xi)^2 - 2 s^{-2} (\partial_\theta \eta) (\partial_\theta \xi) G(\eta) \xi
+ \big( G(\eta) \xi \big)^2 \Big] \,, \\
(\partial_\theta \varphi)^2 \big|_{r=s} & = & \frac{1}{\big( 1 + s^{-2} (\partial_\theta \eta)^2 \big)^2} 
\Big[ (\partial_\theta \xi)^2 + 2 (\partial_\theta \eta) (\partial_\theta \xi) G(\eta) \xi
+ (\partial_\theta \eta)^2 \big( G(\eta) \xi \big)^2 \Big] \,, 
\end{eqnarray*}
yields
\begin{eqnarray*}
(\partial_r \varphi)^2 + s^{-2} (\partial_\theta \varphi)^2 \big|_{r=s}
& = & \frac{1}{\big( 1 + s^{-2} (\partial_\theta \eta)^2 \big)^2} \Big[
s^{-2} \big( 1 + s^{-2} (\partial_\theta \eta)^2 \big) (\partial_\theta \xi)^2
+ \big( 1 + s^{-2} (\partial_\theta \eta)^2 \big) \big( G(\eta) \xi \big)^2 \Big] \,, \\
& = & \frac{1}{1 + s^{-2} (\partial_\theta \eta)^2} \Big[ s^{-2} (\partial_\theta \xi)^2 + \big( G(\eta) \xi \big)^2 \Big] \,.
\end{eqnarray*}
Finally, putting all these expressions together in \eqref{kine}--\eqref{dyna} leads to a closed system of two equations
\begin{eqnarray} \label{ham_kine}
\partial_t \eta & = & -G(\eta) \xi \,, \\
\partial_t \xi & = & -\frac{1}{2 \big( 1 + s^{-2} (\partial_\theta \eta)^2 \big)} \Big[ s^{-2} (\partial_\theta \xi)^2 
+ 2 s^{-2} (\partial_\theta \eta) (\partial_\theta \xi) G(\eta) \xi - \big( G(\eta) \xi \big)^2 \Big] \nonumber \\
& & + \frac{\sigma}{\rho} \kappa \,,
\label{ham_dyna}
\end{eqnarray}
for $\eta$ and $\xi$, which gives a more explicit form of \eqref{canonical} and is completely equivalent to
the original nonlinear formulation \eqref{laplace}--\eqref{farfield} (with $\Delta p = 0$).
Recall that, by definition, the solution of the Laplace equation \eqref{laplace} subject to 
the far-field vanishing condition \eqref{farfield} is encoded in the DNO.
We also point out that, in addition to being Hamiltonian, Eqs. \eqref{ham_kine}--\eqref{ham_dyna}
is a lower-dimensional version of \eqref{laplace}--\eqref{farfield} in terms of surface variables alone.
It is thus more appealing for mathematical analysis and numerical simulation.
In this spirit, a numerical model will be proposed here to solve \eqref{ham_kine}--\eqref{ham_dyna}.
Of course, the contribution $\Delta p$ can be added to \eqref{ham_dyna} as in \eqref{dyna},
but then the Hamiltonian structure would not be preserved in general.

\subsection{Variational derivatives of the Hamiltonian}
\label{Variational}

We now apply Benjamin's formalism \cite{b87} together with the DNO introduced above to show the connection between 
the variational derivatives of $H$ in \eqref{canonical} and the lower-dimensional system \eqref{ham_kine}--\eqref{ham_dyna}
that we just derived.
For this purpose, we rewrite the kinetic part of the Hamiltonian \eqref{ener} as
\[
K = \int_{\cal D} \frac{1}{2} |\nabla \varphi|^2 dA = \int_{\cal D} \frac{1}{2} \big[ \nabla \cdot (\varphi \nabla \varphi) - \varphi \nabla^2 \varphi \big] dA \,, 
\]
where the last term vanishes due to \eqref{laplace}.
Then using the divergence theorem, we obtain
\begin{eqnarray*}
2 K & = & \int_{\cal D} \nabla \cdot (\varphi \nabla \varphi) \,  dA = \int_{\cal S} \varphi (\nabla \varphi \cdot {\bf n}) \, d\ell \,, \\
& = & \int_{\cal S} \varphi (\nabla \varphi \cdot {\bf n}) \sqrt{s^2 + (\partial_\theta \eta)^2} \, d\theta
= \int_0^{2\pi} \xi G(\eta) \xi \, s d\theta \,,
\end{eqnarray*}
by virtue of \eqref{dirichlet} and \eqref{DNO}.
Following Benjamin \cite{b87}, a crucial step is to introduce the weighted element $d\mu = \sqrt{{\cal G}/{\cal G}_{11}} d\theta = s d\theta$ 
along some interval ${\cal C}$, where ${\cal G} = r^2$ and ${\cal G}_{11} = 1$ are respectively the determinant 
and first diagonal entry of the metric tensor for polar coordinates, evaluated on ${\cal S}$.
The Hamiltonian then reads
\begin{equation} \label{ener2}
H = K + P = \frac{1}{2} \int_{\cal C} \xi G(\eta) \xi \, d\mu + \frac{\sigma}{\rho} \int_0^{2\pi} \sqrt{s^2 + (\partial_\theta \eta)^2} \, d\theta \,,
\end{equation}
in terms of the DNO and the two conjugate variables $(\eta,\xi)$.
Given such a functional, the variation $\partial_u H$ is defined via the G\^ateaux derivative
\begin{equation} \label{gateaux}
\langle \partial_u H, v \rangle = \frac{d}{d\lambda} H(u + \lambda v) \Big|_{\lambda=0} \,,
\end{equation}
with respect to the inner product $\langle f, h \rangle = \int_{\cal C} f h \, d\mu$ for any functions $u$, $v$ in some Hilbert space.
From the alternate form \eqref{ener2} of $H$ and the definition \eqref{gateaux} of the variational derivative, we deduce
\begin{eqnarray} \label{variation}
\partial_\xi H & = & \partial_\xi K = G(\eta) \xi \,, \\
\partial_\eta K & = & -\frac{1}{2 \big( 1 + s^{-2} (\partial_\theta \eta)^2 \big)} \Big[ s^{-2} (\partial_\theta \xi)^2 
+ 2 s^{-2} (\partial_\theta \eta) (\partial_\theta \xi) G(\eta) \xi - \big( G(\eta) \xi \big)^2 \Big] \,. \nonumber
\end{eqnarray}
Equation \eqref{variation} for $\partial_\xi H$ follows from the self-adjointness of the DNO \cite{fns07,nn04,nr01}.
Establishing the other identity for $\partial_\eta K$ is a lengthy calculation but is closely related to that for the water wave problem.
Therefore, we refer the reader to \cite{cgs21} for more details.

Regarding the capillary component, we note that $\partial_u H$ can be expressed alternatively as
\[
\int_{\cal C} (\partial_u H) v \, d\mu = \langle \partial_u H, v \rangle
= \int_0^{2\pi} (\partial_u H)^* v \, d\theta \,,
\]
which implies that $\partial_u H = s^{-1} (\partial_u H)^*$
where $(\partial_u H)^*$ is the variational derivative \eqref{gateaux} with respect to the inner product
$\langle f, h \rangle = \int_0^{2\pi} f h \, d\theta$.
Applying this idea to the potential part of \eqref{ener2}, we find
\[
\frac{\rho}{\sigma} (\partial_\eta P)^* = \frac{s^3 + 2 s (\partial_\theta \eta)^2 - s^2 \partial_\theta^2 \eta}{\big( s^2 + (\partial_\theta \eta)^2 \big)^{3/2}} \,,
\]
hence $\partial_\eta P = s^{-1} (\partial_\eta P)^* = \sigma \kappa/\rho$ with $\kappa$ as presented in \eqref{curvature}.
Combining these results for $\partial_\eta H = \partial_\eta K + \partial_\eta P$ together with \eqref{variation} for $\partial_\xi H$
shows the equivalence between \eqref{canonical} and \eqref{ham_kine}--\eqref{ham_dyna}.

\subsection{Variational derivatives of the momentum}

For completeness and reference in a subsequent section, we use the same formalism to evaluate
variational derivatives of the angular momentum \eqref{moment} with respect to the surface variables $(\eta,\xi)$.
Similar to $H$, we first rewrite $I$ explicitly in terms of these variables as follows
\[
I = \int_{\cal D} \partial_\theta \varphi \, dA = \int_{\cal D} \nabla(r^2 \varphi) \cdot \nabla \theta \,  dA
= \int_{\cal D} \nabla \cdot (r^2 \varphi \nabla \theta) \, dA \,,
\]
by virtue of the identity
\[
\nabla \cdot (r^2 \varphi \nabla \theta) = \nabla(r^2 \varphi) \cdot \nabla \theta + r^2 \varphi \nabla^2 \theta \,,
\]
together with the fact that $\nabla^2 \theta = 0$. The divergence theorem implies
\begin{eqnarray}
I & = & \int_{\cal S} s^2 \varphi (\nabla \theta \cdot {\bf n}) \, d\ell 
= \int_{\cal S} s^2 \varphi (\nabla \theta \cdot {\bf n}) \sqrt{s^2 + (\partial_\theta \eta)^2} \, d\theta \,, \nonumber \\
& = & \int_0^{2\pi} \xi \partial_\theta \eta \, s d\theta = \int_{\cal C} \xi \partial_\theta \eta \, d\mu \,,
\label{moment2}
\end{eqnarray}
via \eqref{dirichlet} and \eqref{normal}, after recognizing that $\nabla \theta = (0,r^{-1})^\top$.
It is then a straightforward calculation to show that
\begin{equation} \label{momentderiv}
\partial_\xi I = \partial_\theta \eta \,, \quad \partial_\eta I = -\partial_\theta \xi \,, 
\end{equation}
based on the definition \eqref{gateaux}.
For $\partial_\eta I$, because Eq. \eqref{moment2} involves $\partial_\theta \eta$ and also depends on $\eta$ through $s$, 
an integration by parts in $\theta$ is required to obtain its expression in \eqref{momentderiv}.

Likewise, we easily see that
\begin{equation} \label{levelderiv}
\partial_\xi Q = 0 \,, \quad \partial_\eta Q = 1 \,.
\end{equation}

\section{Dirichlet--Neumann operator}
\label{DtNOp}

The question now is how to calculate the DNO given the boundary $\eta$ and Dirichlet data $\xi$ at any time $t$, 
in order to solve \eqref{ham_kine}--\eqref{ham_dyna}.
The approach that we advocate here is of boundary perturbation type and is based on the fact that
the DNO is analytic in $\eta$ under certain (relatively mild) regularity conditions,
as first shown by Coifman and Meyer \cite{cm85}.
Rigorous proofs of this analyticity property can be found in \cite{nn04} and \cite{fns07} 
for the two-dimensional circular and three-dimensional spherical cases, respectively.

The starting point is the Laplace equation \eqref{laplace} in polar coordinates
\[
\partial_r^2 \varphi + r^{-1} \partial_r \varphi + r^{-2} \partial_\theta^2 \varphi = 0 \,,
\]
for which we consider two different elementary solutions
\begin{equation} \label{harmonic}
\varphi_1 = \ln r \,, \quad \varphi_2 = r^{-n} e^{{\rm i} n \theta} \,,
\end{equation}
with $n \in \mathbb{N}$ (non-negative integers).
The time dependence is omitted because the Laplace equation is solved in a frozen domain at each time $t$.
The first solution $\varphi_1$ describes bubble motion in the $r$-direction alone (i.e. purely circular compression or dilatation)
while the second solution $\varphi_2$ allows for wave development on the bubble surface (i.e. shape distortions) in the $\theta$-direction.
Both phenomena are physically relevant, hence the importance to examine both solutions.
In particular, the radial motion is inherent to this physical problem (even in the absence of waves on the bubble surface)
owing to surface tension associated with the bubble curvature.

Note that both $\varphi_1$ and $\varphi_2$ satisfy the periodic boundary conditions in $\theta$
as well as the vanishing condition \eqref{farfield} as $r \to +\infty$.
This explains why the option $\varphi_2$ with $n < 0$ is ruled out.
There is no singularity at $r = 0$ for either $\varphi_1$ or $\varphi_2$ because this location is outside of the fluid domain ${\cal D}$.

Next we exploit the analyticity property of the DNO by seeking a Taylor series representation in $\eta$, namely
\begin{equation} \label{series}
G(\eta) = \sum_{j=0}^{+\infty} G_j(\eta) \,,
\end{equation}
about the reference circular geometry $\eta = 0$, i.e. $r = R$ with $R$ being a constant.
By construction, each term $G_j$ in \eqref{series} is homogeneous of degree $j$ in $\eta$
and its action as an operator on any basis function $e^{{\rm i} n \theta}$ 
(associated with Fourier mode or wavenumber $n$) can be determined in a recursive manner.
This derivation of the DNO relies on the choice of harmonic function (i.e. solution to the Laplace equation)
and, because $\varphi_1$ and $\varphi_2$ are two different harmonic functions, we will present details for each case separately.
We will first consider $\varphi_2$ which has both $(r,\theta)$-dependences, 
before inspecting the univariate case with $\varphi_1$.
Incidentally, we recognize that $\varphi_2$ should be specified as a real-valued function (corresponding to a velocity potential).
However, its complex-valued form in \eqref{harmonic} is more convenient for the purposes of our derivation
and, because the DNO is a linear operator in $\xi$, it is sufficient to write $e^{{\rm i} n \theta}$.
Contributions from its complex conjugate $e^{-{\rm i} n \theta}$ would be redundant.

Substituting \eqref{harmonic} and \eqref{series} for $\varphi_2$ into \eqref{DNO2} yields
\begin{eqnarray*}
\Big( \sum_{j=0}^{+\infty} G_j(\eta) \Big) (R + \eta)^{-|n|} e^{{\rm i} n \theta}
& = & \Big[ (R + \eta)^{-|n|-1} |n| + {\rm i} (\partial_\theta \eta) (R + \eta)^{-|n|-2} n \Big] e^{{\rm i} n \theta} \,, \\
\Big( \sum_{j=0}^{+\infty} G_j(\eta) \Big) \Big( 1 + \frac{\eta}{R} \Big)^{-|n|} e^{{\rm i} n \theta}
& = & R^{-1} \Big[ \Big( 1 + \frac{\eta}{R} \Big)^{-|n|-1} |n| - \Big( D \frac{\eta}{R} \Big) \Big( 1 + \frac{\eta}{R} \Big)^{-|n|-2} n \Big] e^{{\rm i} n \theta} \,,
\end{eqnarray*}
where $D = -{\rm i} \, \partial_\theta$ and the absolute value $|n|$ is employed to enforce exponent positivity 
for $n \in \mathbb{Z}$ (all integers over the full spectrum), as motivated above.
In terms of the binomial expansion
\[
\Big( 1 + \frac{\eta}{R} \Big)^{-n} = \sum_{\ell=0}^{+\infty} \Big( \frac{\eta}{R} \Big)^\ell C_\ell^{-n} \,,
\]
with coefficient
\[
C_\ell^{-n} = \begin{pmatrix}
-n \\
\ell
\end{pmatrix} = (-1)^\ell \begin{pmatrix}
n + \ell - 1 \\
\ell
\end{pmatrix} = (-1)^\ell \, \frac{n (n + 1) (n + 2) \cdots (n + \ell - 1)}{\ell !} \,,
\]
this equation becomes
\[
\Big( \sum_{j=0}^{+\infty} G_j(\eta) \Big) \Big( \sum_{\ell=0}^{+\infty} \Big( \frac{\eta}{R} \Big)^\ell C_\ell^{-|n|} \Big) e^{{\rm i} n \theta} 
= R^{-1} \Big[ \sum_{\ell=0}^{+\infty} \Big( \frac{\eta}{R} \Big)^\ell C_\ell^{-|n|-1}  |n| - \Big( D \frac{\eta}{R} \Big) 
\sum_{\ell=0}^{+\infty} \Big( \frac{\eta}{R} \Big)^\ell C_\ell^{-|n|-2} n \Big] e^{{\rm i} n \theta} \,.
\]
By inspection, the first term for $j = 0$ in \eqref{series} is given by
\[
G_0 e^{{\rm i} n \theta}  = \frac{|n|}{R} e^{{\rm i} n \theta} \,,
\]
which can be viewed as the Fourier symbol of the pseudo-differential operator
\begin{equation} \label{G0}
G_0 = \frac{|D|}{R} \,,
\end{equation}
acting on any function that has a Fourier series decomposition in $\theta$.
Recall that, by definition, the Fourier symbol associated with $D$ is $n$.
For $j > 0$, collecting terms of the same degree in $\eta$ leads to
\begin{eqnarray*}
G_j(\eta) e^{{\rm i} n \theta} & = & -\sum_{\ell=0}^{j-1} G_\ell(\eta) \Big( \frac{\eta}{R} \Big)^{j-\ell} C_{j-\ell}^{-|n|} e^{{\rm i} n \theta} \\
& & + R^{-1} \Big[ \Big( \frac{\eta}{R} \Big)^j C_j^{-|n|-1}  |n| - \Big( \frac{\eta}{R} \Big)^{j-1} 
\Big( D \frac{\eta}{R} \Big) C_{j-1}^{-|n|-2} n \Big] e^{{\rm i} n \theta} \,,
\end{eqnarray*}
which can also be expressed symbolically as
\begin{eqnarray*}
G_j(\eta) & = & -\sum_{\ell=0}^{j-1} G_\ell(\eta) \Big( \frac{\eta}{R} \Big)^{j-\ell} C_{j-\ell}^{-|D|} \\
& & + R^{-1} \Big[ \Big( \frac{\eta}{R} \Big)^j C_j^{-|D|-1}  |D| - \Big( \frac{\eta}{R} \Big)^{j-1} 
\Big( D \frac{\eta}{R} \Big) C_{j-1}^{-|D|-2} D \Big] \,.
\end{eqnarray*}
By convention, unless parentheses are specified, any operator acts on all functions to its right in the same term.
Pseudo-differential operators like $D$, $|D|$ or $C_j^{-|D|}$ are also called Fourier multipliers
due to their multiplicative action in the Fourier space (dual to the physical $\theta$-space).
Then recognizing that
\[
D \Big( \frac{\eta}{R} \Big)^j D = j \Big( \frac{\eta}{R} \Big)^{j-1} \Big( D \frac{\eta}{R} \Big) D
+ \Big( \frac{\eta}{R} \Big)^j D^2 \,, \quad D^2 = |D|^2 \,,
\]
together with the property
\[
C_j^{-|D|-1} = -C_{j-1}^{-|D|-2} \frac{|D| + 1}{j} = -\frac{1}{j} C_{j-1}^{-|D|-2} |D| - \frac{1}{j} C_{j-1}^{-|D|-2} \,,
\]
for the binomial coefficient, we arrive at
\[
G_j(\eta) = -\frac{1}{j} \Big[ D \Big( \frac{\eta}{R} \Big)^j \frac{D}{R} + \Big( \frac{\eta}{R} \Big)^j G_0 \Big] C_{j-1}^{-|D|-2}
- \sum_{\ell=0}^{j-1} G_\ell(\eta) \Big( \frac{\eta}{R} \Big)^{j-\ell} C_{j-\ell}^{-|D|} \,.
\]
Finally, we invoke the self-adjointness of the DNO to obtain
\begin{equation} \label{Gj}
G_j(\eta) = -\frac{1}{j} \Big[ D \Big( \frac{\eta}{R} \Big)^j \frac{D}{R} + \Big( \frac{\eta}{R} \Big)^j G_0 \Big] C_{j-1}^{-|D|-2}
- \sum_{\ell=0}^{j-1} C_{j-\ell}^{-|D|} \Big( \frac{\eta}{R} \Big)^{j-\ell} G_\ell(\eta) \,,
\end{equation}
after reversing the sequence of application for the various operations in the summation above \cite{fns07,nn04,nr01}.
The reason for doing so will be made more clear when discussing the numerical scheme in a subsequent section.

We now repeat this calculation by using \eqref{harmonic} with $\varphi_1$.
For this purpose, we slightly tweak the harmonic function as follows $\varphi_1 = \ln(R_m/r)$
where the constant radius $R_m$ is chosen sufficiently large such that $R_m \gg R$.
Because $\varphi_1$ only depends on $r$, it  may be associated with the zeroth Fourier mode $n = 0$ in $\theta$.
Accordingly, Eq. \eqref{DNO2} reduces to
\begin{eqnarray*}
\Big( \sum_{j=0}^{+\infty} G_j(\eta) \Big) \mathbb{P}_0 \ln \left( \frac{R_m}{R + \eta} \right) e^{{\rm i} n \theta}
& = & \mathbb{P}_0 \frac{1}{R + \eta} e^{{\rm i} n \theta} \,, \\
\Big( \sum_{j=0}^{+\infty} G_j(\eta) \Big) \mathbb{P}_0 \left[ \ln \left( \frac{R_m}{R} \right) 
- \ln \left( 1 + \frac{\eta}{R} \right) \right] e^{{\rm i} n \theta}
& = & R^{-1} \mathbb{P}_0 \left( 1 + \frac{\eta}{R} \right)^{-1} e^{{\rm i} n \theta} \,,
\end{eqnarray*}
where $\mathbb{P}_0$ denotes the projection onto the zeroth mode $n = 0$, i.e. the DNO only affects this specific mode here.
Then Taylor expanding about $\eta = 0$ gives
\[
\Big( \sum_{j=0}^{+\infty} G_j(\eta) \Big) \mathbb{P}_0 \left[ \ln \left( \frac{R_m}{R} \right) 
- \sum_{\ell=1}^{+\infty} \frac{(-1)^{\ell+1}}{\ell} \Big( \frac{\eta}{R} \Big)^{\ell} \right] e^{{\rm i} n \theta}
= R^{-1} \mathbb{P}_0 \sum_{j=0}^{+\infty} (-1)^j \Big( \frac{\eta}{R} \Big)^j e^{{\rm i} n \theta} \,.
\]
For $j = 0$, we readily infer
\begin{eqnarray*}
G_0 \mathbb{P}_0 \ln \left( \frac{R_m}{R} \right) e^{{\rm i} n \theta} & = & R^{-1} \mathbb{P}_0 e^{{\rm i} n \theta} \,, \\
G_0 \mathbb{P}_0 e^{{\rm i} n \theta} & = & \frac{1}{R \ln(R_m/R)} \mathbb{P}_0 e^{{\rm i} n \theta} \,,
\end{eqnarray*}
which can be written symbolically as 
\begin{equation} \label{G0p}
G_0 \mathbb{P}_0 = \frac{1}{R \ln(R_m/R)} \mathbb{P}_0 \,.
\end{equation}
For $j > 0$, by identifying terms of the same degree in $\eta$, we get
\[
G_j(\eta) \mathbb{P}_0 e^{{\rm i} n \theta} = \frac{1}{R \ln(R_m/R)} \mathbb{P}_0 (-1)^j \Big( \frac{\eta}{R} \Big)^j e^{{\rm i} n \theta}
+ \frac{1}{\ln(R_m/R)} \sum_{\ell=0}^{j-1} G_\ell(\eta) \mathbb{P}_0 \frac{(-1)^{j-\ell+1}}{j-\ell} \Big( \frac{\eta}{R} \Big)^{j-\ell} e^{{\rm i} n \theta} \,,
\]
and thus we can define its action symbolically as
\begin{equation} \label{Gjp}
G_j(\eta) \mathbb{P}_0 = \frac{1}{R \ln(R_m/R)} \mathbb{P}_0 (-1)^j \Big( \frac{\eta}{R} \Big)^j 
+ \frac{1}{\ln(R_m/R)} \sum_{\ell=0}^{j-1} \mathbb{P}_0 \frac{(-1)^{j-\ell+1}}{j-\ell} \Big( \frac{\eta}{R} \Big)^{j-\ell} G_\ell(\eta) \,,
\end{equation}
after invoking again the self-adjointness property (to reverse the sequence of operations in the inner summation).
It now makes sense why $R_m \gg R$ was introduced in such a way, otherwise we would end up with
a possible logarithmic singularity when $R_m = R$ 
(or equivalently when $R = 1$ in some dimensional or dimensionless units for the choice $\varphi_1 = \ln r$).
The range $R_m \gg R$ is preferable over $0 < R_m \ll R$ because it is inside the fluid domain ${\cal D}$
and thus complies with the mathematical formulation \eqref{laplace}--\eqref{farfield} for the fluid flow.
Moreover, in that upper range, $\ln(R_m/R)$ is positive and its variation is relatively mild.

Equations \eqref{G0}--\eqref{Gj} or \eqref{G0p}--\eqref{Gjp} provide recursion formulas to evaluate the DNO
in its series form \eqref{series} given $\eta$ and $\xi$.
Depending on the physical situation under consideration, each set may be used separately,
i.e. either formulas \eqref{G0}--\eqref{Gj} or \eqref{G0p}--\eqref{Gjp}.
More generally, they may be combined by superposition as follows.
The zeroth-order operator ($j = 0$) takes the form
\begin{equation} \label{G0t}
G_0 = G_0^{(1)} + G_0^{(2)} = \frac{|D|}{R} + \frac{1}{R \ln(R_m/R)} \mathbb{P}_0 \,,
\end{equation}
from \eqref{G0} and \eqref{G0p}, while the higher-order contributions ($j > 0$) read
\begin{equation} \label{Gjt}
G_j(\eta) = G_j^{(1)}(\eta) + G_j^{(2)}(\eta) \,,
\end{equation}
with $G_j^{(1)}$ and $G_j^{(2)}$ determined by \eqref{Gj} and \eqref{Gjp} respectively.
This superposition enables the coupling between radial pulsations and shape distortions.
Note in particular how $G_0^{(1)}$ and $G_0^{(2)}$ complement each other in \eqref{G0t} 
over the full spectrum $n \in \mathbb{Z}$, considering that $G_0^{(1)} = |D|/R$ has a trivial effect on the zeroth mode $n = 0$,
while $G_0^{(2)} = R^{-1} \mathbb{P}_0/\ln(R_m/R)$ has a nontrivial effect there.
Compared to the water wave problem in the perturbed rectangular geometry where the DNO has been commonly involved \cite{g19},
the additional component $G_j^{(2)}$ represents a distinctive new feature of the present formulation in polar coordinates.

\section{Approximate regimes}

Several analytical approximations can be made in order to simplify the full nonlinear problem.

\subsection{Linearized problem}
\label{Linearized}

We first investigate the linearized problem for small-amplitude deformations about $\eta = 0$, 
relative to the equilibrium state $r = R$ (with $R$ fixed).
As can be deduced from \eqref{dyna}, by setting all derivatives to zero, this equilibrium corresponds to
\begin{equation} \label{Young}
\frac{\sigma}{R} + \Delta p = 0 \,,
\end{equation}
in light of the expansion 
\[
\kappa = \frac{1}{R} - \frac{1}{R^2} (\partial_\theta^2 \eta + \eta) + O(\eta^2) \,.
\]
Equation \eqref{Young}, also known as the Young--Laplace equation, states that the pressure jump $\Delta p$ 
across the bubble boundary is balanced by the constant curvature due to surface tension.
In this case, Eqs. \eqref{laplace}--\eqref{farfield} via their lower-dimensional form \eqref{ham_kine}--\eqref{ham_dyna} simplify to
\begin{equation} \label{linear}
\partial_t \eta = -G_0 \xi \,, \quad \partial_t \xi = -\frac{\sigma}{\rho R^2} (\partial_\theta^2 \eta + \eta) \,,
\end{equation}
with only contributions from up to first order in $(\eta,\xi)$, which can be combined as
\begin{equation} \label{linear2}
\partial_t^2 \eta - \frac{\sigma}{\rho R^2} G_0 (\partial_\theta^2 \eta + \eta) = 0 \,,
\end{equation}
and similarly for $\xi$.
Having rotating wave solutions of the form $\eta, \xi \sim e^{{\rm i} (n \theta - \omega_0 t)}$ in mind here,
it is sufficient to choose $G_0 = G_0^{(1)}$ from \eqref{G0}.
Equation \eqref{linear2} then implies the linear dispersion relation
\begin{equation} \label{dispersion}
\omega_0^2 = \frac{\sigma}{\rho R^3} |n| (n^2 - 1) \,,
\end{equation}
between the angular frequency $\omega_0$ and wavenumber $n$.
For all $n \in \mathbb{Z}$, we see that $\omega_0^2 \ge 0$ and thus $\omega_0$ is real.
Assuming $n$, $\omega_0 > 0$, the angular phase speed can be derived from \eqref{dispersion} as
\begin{equation} \label{linspeed}
c_0 = \frac{\omega_0}{n} = \sqrt{\frac{\sigma(n^2 - 1)}{\rho R^3 n}} \,.
\end{equation}
Dyachenko \cite{d21} obtained the same result using a different mathematical formulation of this problem.
The modes $n = \pm 1$ would correspond to a rigid translation of the bubble as in the three-dimensional case \cite{tb83},
but this mechanism is ignored here.

\subsection{Second-order Stokes approximation for rotating waves}
\label{Stokes}

Extending this linear analysis to the weakly nonlinear regime, we focus our attention on wave solutions of the form
\begin{equation} \label{traveling}
\eta(\theta,t) = \eta(\Theta) \,, \quad \xi(\theta,t) = \xi(\Theta) \,, \quad \Theta = \theta - c \, t \,,
\end{equation}
which rotate counter-clockwise at constant angular speed $c$ under static pressure \eqref{Young}.
Equations \eqref{ham_kine}--\eqref{ham_dyna} then reduce to the nonlinear system of ordinary differential equations (ODEs)
\begin{eqnarray} \label{kine3}
0 & = & c \, \eta' - G(\eta) \xi \,, \\
0 & = & c \, \xi' - \frac{1}{2 (1 + s^{-2} \eta'^2)} \Big[ s^{-2} \xi'^2 
+ 2 s^{-2} \eta' \xi' G(\eta) \xi - \big( G(\eta) \xi \big)^2 \Big] \nonumber \\
& & + \frac{\sigma}{\rho} \Big( \kappa - \frac{1}{R} \Big) \,,
\label{dyna3}
\end{eqnarray}
by the chain rule, where the primes denote differentiation with respect to $\Theta$.
Similarly, Fourier multipliers with respect to $\theta$ in $G(\eta)$ are replaced by their counterparts with respect to $\Theta$.
The change of variables \eqref{traveling} is equivalent to reformulating this problem in a reference frame rotating at constant speed $c$.

Alternatively, from the Hamiltonian viewpoint, such rotating solutions can be interpreted as fixed points of
the renormalized Hamiltonian $\widehat H = H - c \, I - \sigma Q/(\rho R)$
where $Q$ (mean surface level) and $I$ (angular momentum) are also invariants of motion as defined by \eqref{level} and \eqref{moment}.
In other words, these solutions satisfy
\[
\begin{pmatrix}
0 & -1 \\
1 & 0
\end{pmatrix}
\begin{pmatrix}
\partial_\eta \widehat H \\
\partial_\xi \widehat H
\end{pmatrix} = \begin{pmatrix}
0 \\
0
\end{pmatrix} \,,
\]
which coincide with \eqref{kine3}--\eqref{dyna3}.
This equivalence can be checked directly by invoking \eqref{variation}, \eqref{momentderiv} and \eqref{levelderiv},
modulo the substitution of $\Theta$ for $\theta$.

Along the lines of Stokes theory for the water wave problem \cite{f85}, we seek $(\eta,\xi)$ perturbatively via an asymptotic series
\begin{equation} \label{asymptotic}
\eta = \varepsilon \eta_0 + \varepsilon^2 \eta_1 + \cdots \,, \quad
\xi = \varepsilon \xi_0 + \varepsilon^2 \xi_1 + \cdots \,,
\end{equation}
together with $c = c_0 + \varepsilon c_1 + \cdots$, where the perturbation parameter $\varepsilon \ll 1$
is a dimensionless measure of the wave amplitude (e.g. the wave steepness).
Inserting \eqref{asymptotic} in \eqref{kine3}--\eqref{dyna3} and equating coefficients in factor of the same power in $\varepsilon$,
we find 
\[
c_0 \eta_0' - G_0 \xi_0 = 0 \,, \quad c_0 \xi_0' - \frac{\sigma}{\rho R^2} (\eta_0'' + \eta_0) = 0 \,,
\]
for $(\eta_0,\xi_0)$ at first order $O(\varepsilon)$, which is nothing but the homogeneous linear system \eqref{linear} in terms of $\Theta$
with $c_0$ given by \eqref{linspeed}.
We may take a general solution of the form
\begin{equation} \label{stokes1}
\eta_0 = a_0 \cos(n \Theta) \,, \quad \xi_0 = -a_0 c_0 R \sin(n \Theta) \,, \quad n > 1 \,,
\end{equation}
which sets the amplitude and phase of this Stokes wave.
As stated in the previous section, only the contributions \eqref{G0}--\eqref{Gj} are implemented here for the DNO.
At second order $O(\varepsilon^2)$, we obtain the inhomogeneous linear system 
\begin{eqnarray} \label{eq_xi1}
c_0 \eta_1' - G_0 \xi_1 & = & -c_1 \eta_0' + G_1(\eta_0) \xi_0 \,, \\
c_0 \xi_1' - \frac{\sigma}{\rho R^2} (\eta_1'' + \eta_1) & = & -c_1 \xi_0' + \frac{1}{2} \Big[ \frac{1}{R^2} \xi_0'^2 - (G_0 \xi_0)^2 \Big] 
\nonumber \\
& & -\frac{\sigma}{\rho R^3} \Big( \eta_0^2 + 2 \eta_0 \eta_0'' + \frac{1}{2} \eta_0'^2 \Big) \,, \nonumber 
\end{eqnarray}
for $(\eta_1,\xi_1)$ with inhomogeneous terms depending on $(\eta_0,\xi_0)$.
This system can be merged into the single equation
\begin{eqnarray*}
\left( 1 - \frac{\sigma G_0}{c_0^2 \rho R^2} \right) \eta_1'' - \frac{\sigma G_0}{c_0^2 \rho R^2} \eta_1
& = & -\frac{c_1}{c_0} \eta_0'' + \frac{G_1(\eta_0')}{c_0} \xi_0 + \frac{G_1(\eta_0)}{c_0} \xi_0' \\
& & - \frac{G_0}{c_0^2} \left[ c_1 \xi_0' - \frac{1}{2 R^2} \xi_0'^2 + \frac{1}{2} (G_0 \xi_0)^2 \right. \\
& & \left. + \frac{\sigma}{\rho R^3} \Big( \eta_0^2 + 2 \eta_0 \eta_0'' + \frac{1}{2} \eta_0'^2 \Big) \right] \,,
\end{eqnarray*}
for $\eta_1$ alone after eliminating $\xi_1$.
Note that
\[
G_0 = \frac{|D|}{R} \,, \quad G_1(\eta) = |D| \frac{\eta}{R} G_0 - D \frac{\eta}{R^2} D - \frac{\eta}{R} G_0 \,, 
\]
according to \eqref{G0}--\eqref{Gj}.
Substitution of \eqref{stokes1} for $(\eta_0,\xi_0)$ leads to
\begin{eqnarray}
\left( 1 - \frac{\sigma G_0}{c_0^2 \rho R^2} \right) \eta_1'' - \frac{\sigma G_0}{c_0^2 \rho R^2} \eta_1
& = & \frac{2 c_1 a_0 n^2}{c_0} \cos(n \Theta) + \frac{a_0^2 n^2}{R} \cos(2 n \Theta) + \frac{a_0^2 n^3}{R} \cos(2 n \Theta) \nonumber \\
& & + \frac{5 a_0^2 n^3 \sigma}{2 c_0^2 \rho R^4} \cos(2 n \Theta) - \frac{a_0^2 n \sigma}{c_0^2 \rho R^4} \cos(2 n \Theta) \,.
\label{eq_eta1}
\end{eqnarray}
In doing so, we have applied the identities
\[
D \begin{pmatrix}
\cos(n \Theta) \\
\sin(n \Theta)
\end{pmatrix} = {\rm i} \, n \begin{pmatrix}
\sin(n \Theta) \\
-\cos(n \Theta)
\end{pmatrix} \,, \quad
|D| \begin{pmatrix}
\cos(n \Theta) \\
\sin(n \Theta)
\end{pmatrix} = n \begin{pmatrix}
\cos(n \Theta) \\
\sin(n \Theta)
\end{pmatrix} \,,
\]
for $n > 0$.
Because the second-order corrections are supposed to be bound to the first-order components,
we choose the particular form
\[
\eta_1 = a_1 \cos(2 n \Theta) \,,
\]
as suggested by the right-hand side of \eqref{eq_eta1}.
Then equating the coefficients in factor of $\cos(n \Theta)$ and $\cos(2 n \Theta)$ implies $c_1 = 0$ for $a_0 \neq 0$ 
(nontrivial solutions) and
\[
a_1 = \frac{a_0^2 (2 n^3 + 7 n^2 - 2 n - 4)}{4 R (2 n^2 + 1)} \,,
\]
respectively. Once $\eta_1$ is known, $\xi_1$ can be determined from \eqref{eq_xi1}, namely
\begin{equation} \label{eq2_xi1}
G_0 \xi_1 = c_0 \eta_1' - G_1(\eta_0) \xi_0 = -c_0 n \left( 2 a_1 + \frac{a_0^2}{2 R} \right) \sin(2 n \Theta) \,,
\end{equation}
which yields
\[
\xi_1 = -c_0 \left( a_1 R + \frac{a_0^2}{4} \right) \sin(2 n \Theta) \,,
\]
as dictated by the right-hand side of \eqref{eq2_xi1}.
In principle, this procedure can be pursued up to an arbitrary order at the expense that the equations become 
increasingly more complicated (involving higher-order terms $G_j$) as the level of approximation rises.
Collecting all these expressions in \eqref{asymptotic} given $a_0$ and $n$ provides an explicit weakly nonlinear estimate,
accurate up to second order in Stokes theory, for steadily rotating waves at angular speed $c$.
It may serve as a validation test for two-dimensional numerical solvers including the present one,
when applied to wave speeds near $c_0$.
Numerical illustrations will be shown in a subsequent section.

\subsection{Rayleigh--Plesset model}

The Rayleigh--Plesset (RP) equation and its variants have been an important model to understand the dynamics of cavitation bubbles \cite{fl97,p17}.
For a single bubble in an infinite body of incompressible fluid and under the assumption of spherical symmetry, 
a second-order ODE can be derived for the time evolution of its radius.
Because this model has been mostly considered for three-dimensional spherical applications in the literature,
we find it suitable to present its detailed derivation in the two-dimensional circular configuration 
so as to make this paper sufficiently self-contained.
We also restrict ourselves to the inviscid limit, although the RP equation typically includes a viscous term.

Based on the potential-flow formulation \eqref{laplace}--\eqref{farfield} in the special case with circular symmetry,
we take a harmonic function of the form
\begin{equation} \label{RPharmonic}
\varphi = B \ln \left( \frac{R_m}{r} \right) \,, \quad \partial_r \varphi = -\frac{B}{r} \,,
\end{equation}
as in Sec. \ref{DtNOp}, and define the free surface by ${\cal S} = \{ 0 \le \theta < 2\pi, r = s(t) \}$.
Here we prefer to employ the variable $s$ (full surface deformation) rather than $\eta$ (surface perturbation relative to $r = R$)
in order to comply with the typical formulation of the RP equation.
The kinematic boundary condition \eqref{kine} then simplifies to $ds/dt = \partial_r  \varphi$ on ${\cal S}$,
which implies that $B = - s \, ds/dt$.
The dynamic boundary condition \eqref{dyna} reduces to
\begin{equation} \label{dyna2}
\partial_t \varphi = -\frac{1}{2} (\partial_r \varphi)^2 + \frac{\sigma}{\rho s} + \frac{1}{\rho} \Delta p \,,
\end{equation}
where
\[
\partial_r \varphi = \frac{ds}{dt} \,, \quad
\partial_t \varphi = -\left[ \left( \frac{ds}{dt} \right)^2 + s \frac{d^2 s}{dt^2} \right] \ln \left( \frac{R_m}{s} \right) \,, 
\quad \mbox{on} \quad {\cal S} \,.
\]
To complete this boundary condition, we need to specify the pressure jump $\Delta p = p_\infty - p_B$.
The bubble pressure is assumed to obey the static law
\[
p_B = \left( p_\infty + \frac{\sigma}{R} \right) \left( \frac{V_0}{V} \right)^\gamma \,, \quad \gamma \geq 0 \,,
\]
for a polytropic process so that it satisfies \eqref{Young} at equilibrium $r = R$, 
where the circular bubble volume is given by $V = \pi s^2$ at any time $t$ according to \eqref{mass2} 
and $V_0 = \pi R^2$ denotes its equilibrium value \cite{lh89}.
With all these substitutions, Eq. \eqref{dyna2} becomes
\begin{equation} \label{RPeq}
s \frac{d^2 s}{dt^2} \ln \left( \frac{R_m}{s} \right) + \left( \frac{ds}{dt} \right)^2 \left[ \ln \left( \frac{R_m}{s} \right) - \frac{1}{2} \right]
+ \frac{\sigma}{\rho s} + \frac{p_\infty}{\rho} - \frac{1}{\rho} \left( p_\infty 
+ \frac{\sigma}{R} \right) \left( \frac{R}{s} \right)^{2\gamma} = 0 \,.
\end{equation}
This two-dimensional version of the RP equation is somewhat different from its typical three-dimensional counterpart,
due to the different forms of harmonic function between these two geometries.

In three dimensions and in the absence of viscosity, the RP equation is known to possess a canonical Hamiltonian structure 
assuming a uniform far-field pressure $p_\infty$ \cite{fl97}.
Such a mathematical formulation is suitable for a phase-plane stability analysis of solutions about fixed points
when varying the parameter $p_\infty$ \cite{mw62}.
We now reveal a similar Hamiltonian structure for the two-dimensional model \eqref{RPeq}
by following Feng and Leal \cite{fl97} with a key adjustment of their change of variables.
If we set
\begin{equation} \label{RPcoord}
\mathrm{q} = \ln \left( \frac{R_m}{s} \right) \,, \quad \mathrm{p} = -s^3 \frac{ds}{dt} \ln \left( \frac{R_m}{s} \right) \,,
\end{equation}
as motivated by \eqref{RPharmonic}, then Eq. \eqref{RPeq} is equivalent to the system of two first-order ODEs
\begin{eqnarray} \label{eq_q}
\frac{d\mathrm{q}}{dt} & = & \frac{\mathrm{p} \, e^{4\mathrm{q}}}{\mathrm{q} \, R_m^4} \,, \\
\frac{d\mathrm{p}}{dt} & = & -\frac{2\mathrm{p}^2 \, e^{4\mathrm{q}}}{\mathrm{q} \, R_m^4}
+ \frac{\mathrm{p}^2 \, e^{4\mathrm{q}}}{2 \mathrm{q}^2 \, R_m^4} + \frac{\sigma R_m}{\rho} e^{-\mathrm{q}} \nonumber \\
& & + \frac{p_\infty R_m^2}{\rho} e^{-2\mathrm{q}}
- \frac{R_m^2}{\rho} \left( p_\infty + \frac{\sigma}{R} \right) \left( \frac{R}{R_m} \right)^{2\gamma} e^{-2(1-\gamma) \mathrm{q}} \,.
\label{eq_p}
\end{eqnarray}
The first equation is simply an identity from the definitions \eqref{RPcoord} of $\mathrm{q}$ and $\mathrm{p}$.
The second equation is an alternate form of \eqref{RPeq} in terms of these auxiliary variables.
Moreover, this system can be expressed as
\[
\frac{d\mathrm{q}}{dt} = \partial_{\mathrm{p}} {\cal H} \,, \quad \frac{d\mathrm{p}}{dt} = -\partial_{\mathrm{q}} {\cal H} \,,
\]
which are canonical Hamiltonian equations for the two conjugate coordinates $\mathrm{q}$ and $\mathrm{p}$,
associated with the Hamiltonian
\begin{equation} \label{RPhamil}
{\cal H} = \frac{\mathrm{p}^2 \, e^{4\mathrm{q}}}{2 q \, R_m^4} + \frac{\sigma R_m}{\rho} e^{-\mathrm{q}}
+ \frac{p_\infty R_m^2}{2 \rho} e^{-2\mathrm{q}}
- \frac{R_m^2}{2\rho (1-\gamma)} \left( p_\infty + \frac{\sigma}{R} \right) \left( \frac{R}{R_m} \right)^{2\gamma} e^{-2(1-\gamma) \mathrm{q}} \,.
\end{equation}
All these relations can be verified by direct calculation given \eqref{RPcoord} and \eqref{RPhamil}.
This Hamiltonian structure for \eqref{RPeq} is of interest in its own right because it is not directly related to
the Hamiltonian formulation \eqref{canonical} of the full system.
It is not a straightforward extension of that in the three-dimensional case due to the presence
of the logarithm function $\ln(R_m/s)$ in this two-dimensional problem.
As a result, the new Hamiltonian \eqref{RPhamil} and associated evolution equations \eqref{eq_q}--\eqref{eq_p}
exhibit both rational and exponential dependences on $q$.

Note that the possible singularity in \eqref{RPhamil} at $\gamma = 1$ for an isothermal process 
is consistent with the fact that such conditions typically occur when a system is in contact with an outside thermal reservoir,
allowing e.g. for heat exchange in order to maintain a constant temperature.
Accordingly, energy conservation is not expected in this system.
On the other hand, for a general isentropic process ($\gamma \neq 1$) where the conditions are adiabatic and reversible,
energy conservation as represented by \eqref{RPhamil} is realizable.

The linearized problem for the RP equation \eqref{RPeq} about equilibrium $r = R$ also requires an examination.
Substituting $s(t) = R + \eta(t)$ into \eqref{RPeq} and retaining terms of up to first order in $\eta$ gives
\begin{equation} \label{RPlinear}
\frac{d^2 \eta}{dt^2} + \Omega_0^2 \, \eta = 0 \,,
\end{equation}
for a harmonic oscillator at frequency 
\begin{equation} \label{RPdisp}
\Omega_0^2 = \frac{2 \gamma \left( p_\infty + \frac{\sigma}{R} \right) - \frac{\sigma}{R}}{\rho R^2 \ln(R_m/R)} \,,
\end{equation}
if the far-field pressure $p_\infty$ is assumed to be uniform.
This expression for the linear fundamental frequency of radial oscillations shares similarities with the three-dimensional version \cite{pp77}.
Not surprisingly, Eqs. \eqref{RPlinear}--\eqref{RPdisp} coincide exactly with \eqref{linear2} for $\gamma = 0$
and under circular symmetry (i.e. $\theta$-invariance) when $G_0 = G_0^{(2)}$ according to \eqref{G0p}.
The fact that $\Omega_0^2 < 0$ in this situation ($\gamma = 0$, $R_m \gg R$)
is indicative of an instability triggering bubble collapse due to surface tension in the absence of internal pressure.

\section{Numerical methods}

We present a numerical scheme to compute the DNO and discretize \eqref{ham_kine}--\eqref{ham_dyna} in space and time.
We emphasize that, while the DNO is given by a Taylor series expansion, the full system \eqref{ham_kine}--\eqref{ham_dyna}
is directly solved for the time evolution problem (as opposed to invoking a reduced asymptotic model).

\subsection{Space discretization}
\label{Spectral}

Considering the periodic boundary conditions in $\theta$, we use a pseudo-spectral method to discretize the DNO
and equations of motion \eqref{ham_kine}--\eqref{ham_dyna} in space \cite{chqz88}.
This is a natural choice for the computation of $G(\eta)$ because each term in its Taylor series \eqref{series}
is evaluated via recursion formulas \eqref{G0t}--\eqref{Gjt} involving concatenations of Fourier multipliers with powers of $\eta/R$.
More specifically, both functions $\eta$ and $\xi$ are expressed as truncated Fourier series
\[
\begin{pmatrix}
\eta_j \\
\xi_j
\end{pmatrix} = \sum_{n=-N/2}^{N/2-1} \begin{pmatrix}
\widehat \eta_n \\
\widehat \xi_n
\end{pmatrix} e^{{\rm i} n \theta_j} \,, \quad \theta_j = \frac{2\pi}{N} j \,, \quad j = 0, \dots, N-1 \,.
\]
Applications of spatial derivatives or Fourier multipliers are performed in the Fourier space,
while nonlinear products are calculated in the physical space on a regular grid of $N$ collocation points.
For example, if we wish to apply the zeroth-order operator $G_0^{(1)}$ to a function $\xi$ in the physical space,
we implement it as follows
\[
G_0^{(1)} \xi = {\cal F}^{-1} \left( \frac{|n|}{R} {\cal F}(\xi) \right) = {\cal F}^{-1} \left( \frac{|n|}{R} \widehat \xi_n \right) \,.
\]
Similarly, the projection $\mathbb{P}_0$ in $G_j^{(2)}$ can be computed as
\[
\mathbb{P}_0 \xi = {\cal F}^{-1} \big( \widehat \xi_0 \big) \,,
\]
where ${\cal F}$ (resp. ${\cal F}^{-1}$) denotes the direct (resp. inverse) Fourier transform.
All operations from the physical to Fourier space and vice versa are carried out via the fast Fourier transform (FFT).

The Taylor series of the DNO is also truncated to a finite number of terms 
\begin{equation} \label{series2}
G(\eta) \simeq G^{(M)}(\eta) = \sum_{j=0}^M G_j(\eta) \,,
\end{equation}
for which the choice of truncation order $M$ will be discussed in more detail in a subsequent section when showing convergence tests.
Due to the analyticity property, a small number of terms would be sufficient (typically $M < 10 \ll N$) to achieve highly accurate results.
For this computation, the adjoint formulas \eqref{Gj} and \eqref{Gjp} are more efficient than the original ones
(those before reversing the sequence of inner operations) because they allow us to save and re-use the $G_\ell$'s on $\xi$ as vectors,
without having to re-compute these operators at each order $j$ when applied to concatenations of Fourier multipliers and powers of $\eta/R$.
In this adjoint form, the computational cost for evaluating \eqref{series2} can be estimated to be $O(M^2 N \ln N)$ via the FFT.
This point highlights another advantage of the present approach where the Taylor series representation of the DNO together with
the recursive calculation of its constitutive terms avoids setting up and solving a large or dense matrix system
(related to the Laplace problem \eqref{laplace}--\eqref{farfield}),
which contrasts with the classical strategy in other numerical solvers such as
boundary integral methods or volumetric finite-difference/element methods \cite{gg06}.

\subsection{Time integration}
\label{Integration}

Time integration of \eqref{ham_kine}--\eqref{ham_dyna} is performed in the Fourier space so that 
linear terms can be solved exactly by the integrating factor technique.
To do so, we first split these equations into 
\begin{equation} \label{system}
\partial_t \bma{v} = {\cal L} \bma{v} + {\cal N}(\bma{v}) \,,
\end{equation} 
for $\bma{v} = (\eta,\xi)^\top$, where the linear part is given by
\begin{equation} \label{Lv}
{\cal L} \bma{v} = \begin{pmatrix}
0 & -G_0^{(1)} \\
-\frac{\sigma}{\rho R^2}(\partial_\theta^2 + 1) & 0
\end{pmatrix} \begin{pmatrix}
\eta \\
\xi
\end{pmatrix} \,,
\end{equation}
according to \eqref{linear}, and the nonlinear part takes the form ${\cal N}(\bma{v}) = ({\cal N}_1,{\cal N}_2)^\top$ with
\begin{eqnarray*}
{\cal N}_1 & = & -\big( G(\eta) - G_0^{(1)} \big) \xi \,, \\
{\cal N}_2 & = & -\frac{1}{2 \big( 1 + s^{-2} (\partial_\theta \eta)^2 \big)} \Big[ s^{-2} (\partial_\theta \xi)^2 
+ 2 s^{-2} (\partial_\theta \eta) (\partial_\theta \xi) G(\eta) \xi - \big( G(\eta) \xi \big)^2 \Big] \nonumber \\
& & + \frac{\sigma}{\rho} \Big[ \kappa + \frac{1}{R^2}(\partial_\theta^2 \eta + \eta) \Big] + \frac{1}{\rho} \Delta p \,.
\end{eqnarray*}
The pressure difference $\Delta p$ is also included in ${\cal N}_2$ to tackle out-of-equilibrium regimes.
The subtraction of $-G_0^{(1)} \xi$ in ${\cal N}_1$ and of $-\sigma (\partial_\theta^2 \eta + \eta)/(\rho R^2)$ in ${\cal N}_2$
is meant to compensate for their presence in the linear part ${\cal L} \bma{v}$.
Curvature terms are known to cause stiffness in the numerical simulation of capillarity-dominated flows \cite{hls94},
hence treating their linear contributions exactly will help mitigate this issue.

More specifically, we take the Fourier transform of \eqref{system} and make the change of variables
\begin{equation} \label{newvar}
\widehat{\bma{v}}_n(t) = \Phi_n(t) \widehat{\bma{w}}_n(t) \,,
\end{equation}
in terms of the integrating factor
\begin{equation} \label{factor}
\Phi_n(t) = \begin{pmatrix}
\cos(\omega_0 t) & -\frac{|n|}{\omega_0 R} \sin(\omega_0 t) \\
\frac{\sigma (n^2 - 1)}{\omega_0 R^2} \sin(\omega_0 t) & \cos(\omega_0 t)
\end{pmatrix} \,,
\end{equation}
where $\omega_0$ is given by \eqref{dispersion}.
This integrating factor is the fundamental matrix of the linear system 
\[
\partial_t \widehat{\bma{v}}_n = \widehat{\cal L}_n \widehat{\bma{v}}_n = \begin{pmatrix}
0 & -\frac{|n|}{R}\\
\frac{\sigma}{\rho R^2}(n^2 - 1) & 0
\end{pmatrix} \begin{pmatrix}
\widehat \eta_n \\
\widehat \xi_n
\end{pmatrix} \,,
\]
associated with the linear part \eqref{Lv}.
It thus encodes the exact linear solution of \eqref{system}.
Because $\omega_0$ vanishes at $n = \{ 0, \pm 1 \}$, there may be some indetermination at these modes
for either one of the off-diagonal entries in $\Phi_n(t)$, as suggested by \eqref{factor}.
Via l'H\^opital's rule, we find
\[
\Phi_0(t) = \begin{pmatrix}
1 & 0 \\
-\frac{\sigma t}{R^2} & 1
\end{pmatrix} \,,
\]
for $n = 0$, and
\[
\Phi_{\pm 1}(t) = \begin{pmatrix}
1 & -\frac{t}{R} \\
0 & 1
\end{pmatrix} \,,
\]
for $n = \pm 1$.

This leads to a strictly nonlinear system 
\[
\partial_t \widehat{\bma{w}}_n = \Phi_n(t)^{-1} \widehat{\cal N}_n \big[ \Phi_n(t) \widehat{\bma{w}}_n \big] \,,
\]
for the new variable $\widehat{\bma{w}}_n$,
which is solved numerically in time using the fourth-order Runge--Kutta method with constant step $\Delta t$.
The resulting scheme can be inverted back to $\widehat{\bma{v}}_n$ by virtue of \eqref{newvar}, yielding
\[
\widehat{\bma{v}}_n^{k+1} = \Phi_n(\Delta t) \widehat{\bma{v}}_n^k + \frac{\Delta t}{6} \Phi_n(\Delta t) (f_1 + 2 f_2 + 2 f_3 + f_4) \,,
\]
for the numerical solution at time $t_{k+1} = t_k + \Delta t$, where
\begin{eqnarray*}
f_1 & = & \widehat{\cal N}_n(\widehat{\bma{v}}_n^k) \,, \\
f_2 & = & \Phi_n \left( -\frac{\Delta t}{2} \right) \widehat{\cal N}_n \left[ \Phi_n \left( \frac{\Delta t}{2} \right) 
\left( \widehat{\bma{v}}_n^k + \frac{\Delta t}{2} f_1 \right) \right] \,, \\
f_3 & = & \Phi_n \left( -\frac{\Delta t}{2} \right) \widehat{\cal N}_n \left[ \Phi_n \left( \frac{\Delta t}{2} \right) 
\left( \widehat{\bma{v}}_n^k + \frac{\Delta t}{2} f_2 \right) \right] \,, \\
f_4 & = & \Phi_n(-\Delta t) \, \widehat{\cal N}_n \Big[ \Phi_n(\Delta t) \big( \widehat{\bma{v}}_n^k + \Delta t \, f_3 \big) \Big] \,.
\end{eqnarray*}
Here we have exploited the fact that $\Phi_n(t)$ is a semigroup that satisfies the properties
\[
\Phi_n(t)^{-1} = \Phi_n(-t) \,, \quad \Phi_n(t + \tau) = \Phi_n(t) \Phi_n(\tau) \,.
\]
These identities can be verified by direct calculation.
Again, in this time-integration process, the FFT makes it possible to go back and forth 
between $\bma{v}$ and $\widehat{\bma{v}}_n$ in an efficient manner.

\subsection{De-aliasing and filtering}
\label{Filtering}

For pseudo-spectral methods applied to nonlinear problems, numerical errors may stem from the aliasing phenomenon \cite{chqz88}.
In the present algorithm, aliasing may occur when evaluating the equations of motion and the DNO with the FFT.
The $j$th-order term $G_j(\eta) \xi$ involves nonlinearities of degree $j + 1$ as indicated by \eqref{Gj} or \eqref{Gjp}, 
therefore aliasing may be severe for large $j$.
To deal with this issue, the zero-padding technique is a simple and effective option.
Typically, for a quadratic nonlinearity and given resolution $N$, this amounts to doubling the size of the discretized spectra of $(\eta,\xi)$
and setting the Fourier coefficients for the extra modes to zero, consistent with the $2/3$ rule for de-aliasing in this configuration \cite{chqz88}.
Because nonlinearities are of polynomial or rational type in \eqref{ham_kine}--\eqref{ham_dyna} and \eqref{series},
we accommodate each nonlinear term by breaking it up into successive products of two functions
and by applying the aforementioned de-aliasing technique at each multiplicative step.
We have successfully employed such a procedure in other physical settings \cite{gp12,gp16,xg09} 
where nonlinear surface waves were simulated.
It comes along with increased memory storage but this turns out not to be a major concern in the present two-dimensional case via the FFT.

For shape distortions of appreciable steepness, we have found it necessary to filter the numerical solution 
in order to stabilize the computation so that it can run over a sufficiently long period of time.
Otherwise, spurious high-wavenumber oscillations tend to develop, eventually leading to the computation breakdown.
Possible causes for this phenomenon include numerical ill-conditioning of the DNO in its series form \eqref{series2}
(which will be further assessed in a subsequent section), or the ill-posed character of the governing equations.
A detailed mathematical analysis on these points is outside the scope of this study.
As a remedy, we apply a hyperviscosity-type filter of the form
\[
{\exp} \left( -36 \left| \frac{n}{N/2} \right|^{36} \right) \,,
\]
to both Fourier coefficients $\widehat \eta_n$ and $\widehat \xi_n$ at each time step,
where $N/2$ is the largest wavenumber resolved by the spatial discretization.
Such a filter has been commonly adopted in numerical simulations of nonlinear fluid flows by spectral methods
\cite{chqz88,hl07,hls94,gn07,xg09}.
Its smooth but steep behavior near $n = N/2$ ensures that only Fourier coefficients at high wavenumbers are affected.
Accordingly, for a sufficiently high resolution, this filtering strategy should help suppress spurious instabilities while preserving the overall solution.
Combined with zero-padding, it contributes further to control of aliasing errors.

\section{Numerical results}

We present numerical results to illustrate the performance of this numerical model.
These include convergence tests on the DNO computation and direct simulations of time-dependent nonlinear solutions.
For the latter, we explore purely circular motions under transient excitation as well as
shape deformations that rotate steadily or oscillate periodically on the bubble surface.

All variables are non-dimensionalized with respect to the characteristic length $R$, mass $\rho R^3$
and time $\sqrt{\rho R^3/\sigma}$ so that $R = 1$, $\rho = 1$ and $\sigma = 1$ in the resulting dimensionless equations.
For convenience, we will use the same notations but it is now understood that 
values of any variable or parameter are dimensionless.

\subsection{Convergence tests on the DNO}
\label{Convergence}

Because we approximate the DNO in terms of a Taylor series as justified by its analyticity property,
it is suitable to assess its numerical convergence which is expected to be exponential with respect to $M$
for sufficiently smooth deformations $\eta$.
This question is especially relevant considering that such a representation has been shown to be ill-conditioned numerically 
in related problems \cite{gp16,nr01,xg09}.
Indeed, the series expansion \eqref{series2} relies on cancellation of terms to guarantee convergence
but in practice an exact cancellation never happens due to round-off errors.
As indicated by \eqref{Gj}, the higher the order $j$ of $G_j$, the stronger the exponentiation power of $|D|$ in such Fourier multipliers as $C_j^{-|D|}$.
Therefore, numerical errors can be amplified dramatically through the recursive process as the order $j$ increases in \eqref{series2}.
Needless to say that this error amplification will contaminate the numerical solution during the time integration 
of \eqref{ham_kine}--\eqref{ham_dyna}, and may precipitate the accuracy deterioration or promote spurious instabilities.

Here time is frozen and Eqs. \eqref{system}--\eqref{Lv} are not solved yet.
Instead, we inspect the numerical convergence of \eqref{series2} by testing this approximation against an exact expression
that can be derived based on the harmonic function $\varphi_2$ from \eqref{harmonic} together with a specific profile of $\eta$.
More precisely, using the real-valued form
\begin{equation} \label{harmonic2}
\varphi_2 = r^{-n} \sin(n \theta) \,, \quad n > 0 \,,
\end{equation}
and substituting it into the definition \eqref{DNO2}, we obtain
\begin{eqnarray} \label{exact}
G^{(E)}(\eta) \xi & = & -\partial_r \varphi_2 + s^{-2} (\partial_\theta \eta) (\partial_\theta \varphi_2) \big|_{r=s} \,, \nonumber \\
& = & n (R + \eta)^{-n-2} \Big[ (R + \eta) \sin(n \theta) + (\partial_\theta \eta) \cos(n \theta) \Big] \,.
\end{eqnarray}
We perform tests for two different surface profiles
\begin{equation} \label{eta1}
\eta = a_1 \cos(n \theta) \,,
\end{equation}
and
\begin{equation} \label{eta2}
\eta = a_2 \left[ \theta^4 (2\pi - \theta)^4 - \frac{128 \pi^8}{315} \right] \,,
\end{equation}
in combination with \eqref{series2} (numerical approximation) or \eqref{exact} (exact expression) for the DNO.
The choice \eqref{eta1} corresponds to a smooth boundary ($C^\infty$ profile) 
while \eqref{eta2} represents a rougher boundary (with finite smoothness) \cite{nr01}.
Both profiles are prescribed in such a way that their mean value is zero.
Note also that, as opposed to \eqref{eta1} which is more symmetric, the profile \eqref{eta2} tends to be more eccentric to the left,
with a more ovate shape, as $a_2$ is increased.
The maximum distortion is $\| \eta \|_\infty = a_1$ (at any crest or trough) for \eqref{eta1},
while it is given by $\| \eta \|_\infty = a_2 187 \pi^8/315$ for \eqref{eta2} and is achieved at $\theta = \pi$ (left side of the bubble).
Recalling that the DNO \eqref{DNO} is linear in $\xi$ but depends nonlinearly on $\eta$, 
any constant coefficient in \eqref{harmonic2} would be superfluous while the amplitudes $a_1$, $a_2$
in \eqref{eta1}, \eqref{eta2} are expected to be relevant.

In all our computations, we set the parameter $R_m = 1000 R$ so that $R_m \gg R$ as mentioned earlier. 
We have checked that results are insensitive to the specific value of $R_m$ in this range.

\begin{figure}
\centering
\subfloat[]{\includegraphics[width=.5\linewidth]{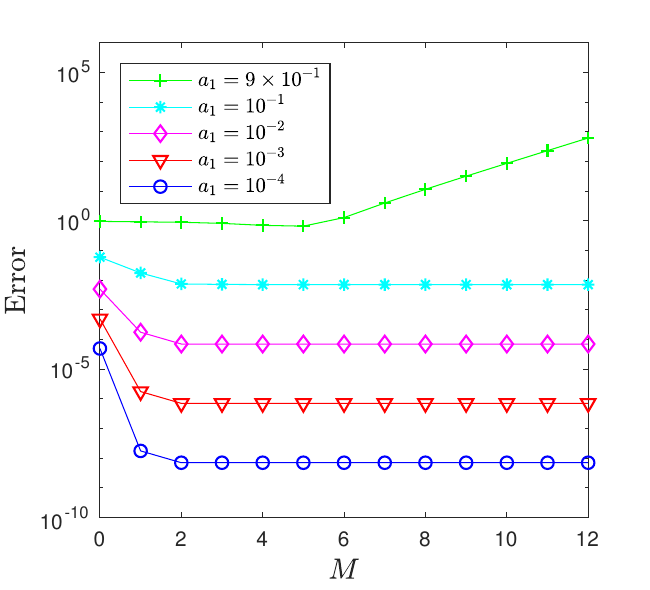}}
\hfill
\subfloat[]{\includegraphics[width=.5\linewidth]{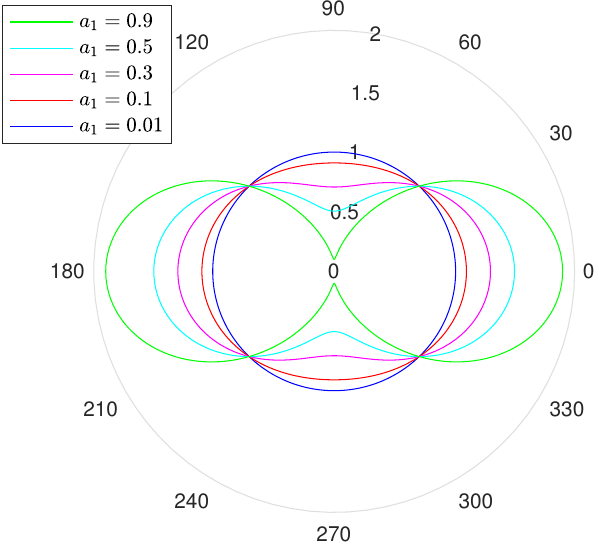}}
\caption{(a) Relative $L^2$ error on the DNO vs. truncation order $M$ for $\eta$ given by \eqref{eta1}.
(b) Surface profile $s$ in polar coordinates $(r,\theta)$. Angles are indicated in degrees.
In each panel, graphs are shown for varying amplitude $a_1$ with $n = 2$ and $N = 256$.}
\label{dno_k2_cos}
\end{figure}

Figure \ref{dno_k2_cos}(a) plots the relative $L^2$ error
\begin{equation} \label{error}
{\rm Error} = \frac{\| G^{(M)}(\eta) \xi - G^{(E)}(\eta) \xi \|_2}{\| G^{(E)}(\eta) \xi \|_2} \,,
\end{equation}
between \eqref{series2} and \eqref{exact} as a function of truncation order $M$ for varying amplitude $a_1$
in the case of a smooth profile \eqref{eta1} with wavenumber $n = 2$ and resolution $N = 256$
(i.e. angular grid size $\Delta \theta = 2\pi/N = 0.024$).
A first observation concerns the convergence with respect to $a_1$ which is demonstrated
by the lower error curves for smaller $a_1$, spanning multiple orders of magnitude.
Given $a_1$, exponential convergence seems to take place over only the first few values of $M$ 
(as indicated by the near-linear slope in these semilog plots). Past $M \simeq 2$, the errors quickly stagnate, 
though they remain small in general.
For large amplitudes, say $a_1 = 0.9$ (which is comparable to the radius $R = 1$ of the unperturbed circular bubble),
the errors are large and reach $100$\%.
After some stagnation, we see a dramatic error growth past some critical value of $M$ (past $M \simeq 5$ for $a_1 = 0.9$).
This phenomenon is characteristic of the numerical ill-conditioning for the DNO computation via its series form \eqref{series2},
as reported by previous studies on free-surface flows in different geometric configurations \cite{g17,gp16,nr01,xg09}.
To help the reader visualize the present geometry, Fig. \ref{dno_k2_cos}(b) portrays various possible bubble shapes 
when varying $a_1$.
For $a_1 = 0.9$ (with $n = 2$), the bubble is so strongly pinched along the central vertical axis that
two lobes are generated.
The surface profile $\eta$ is clearly an important factor influencing the accuracy of the DNO computation,
as pointed out earlier, and this is further illustrated in our next tests.

The fact that convergence promptly stagnates with respect to $M$ is likely another outcome of this numerical ill-conditioning.
A similar phenomenon has been observed in the axisymmetric cylindrical case for surface deformations of a ferrofluid jet \cite{gp16},
but it differs somewhat from results in the rectangular geometry with Cartesian coordinates \cite{g17,nr01,xg09}
where faster (i.e. exponential) convergence is achieved over several values of $M$ before a sudden deterioration occurs.
This difference may be explained here by the action of binomial-type Fourier multipliers like $C_j^{-|D|}$
(aside from $G_0$), which are more complicated than the mere power-type $|D|^j$ in the Cartesian coordinate setting.
As a consequence, the present recursion formula \eqref{Gj} for each $G_j$ contains more nonlinear terms 
(without counting contributions from \eqref{G0p}), and their number increases with $j$.
This promotes stronger amplification of numerical errors, even at low values of $M$.

\begin{figure}
\centering
\subfloat[]{\includegraphics[width=.5\linewidth]{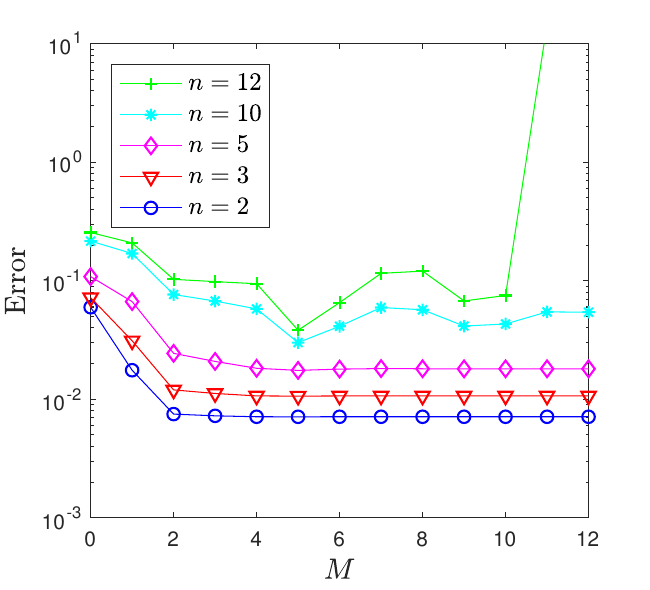}}
\hfill
\subfloat[]{\includegraphics[width=.5\linewidth]{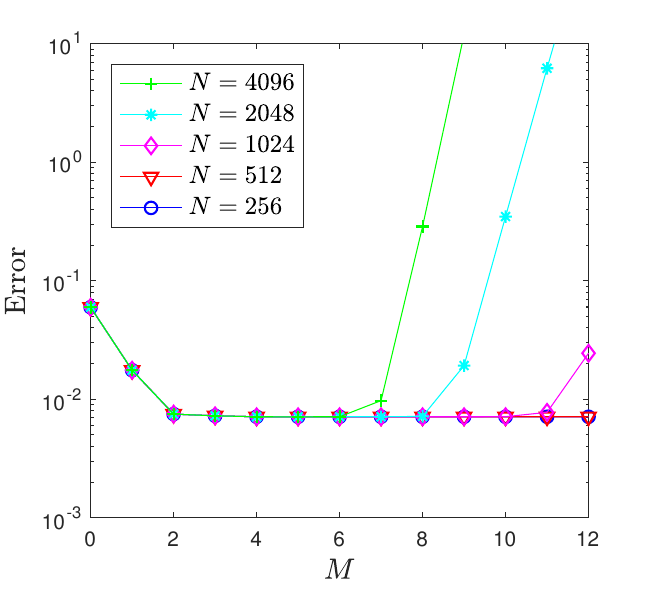}}
\caption{Relative $L^2$ error on the DNO vs. truncation order $M$ for $\eta$ given by \eqref{eta1}.
Graphs are shown for (a) varying wavenumber $n$ (with $a_1 = 0.1$, $N = 256$) 
and (b) varying resolution $N$ (with $a_1 = 0.1$, $n = 2$).}
\label{dno_a01_cos}
\end{figure}

Figure \ref{dno_a01_cos} shows $L^2$ errors \eqref{error} versus $M$ for varying $n$ (with $a_1 = 0.1$, $N = 256$)
and varying $N$ (with $a_1 = 0.1$, $n = 2$).
Here a moderately large amplitude $a_1 = 0.1$ is selected as a compromise
so we can expect small errors while possibly capturing ill-conditioning effects.
In fact, quite large surface steepnesses $\varepsilon = (a_1 n)/R$ are examined in Fig. \ref{dno_a01_cos}(a),
e.g. $\varepsilon = \{ 0.5, 1.0, 1.2 \}$ for $n = \{ 5, 10, 12 \}$ respectively.
Overall, we discern similar features to Fig. \ref{dno_k2_cos}.
The higher the steepness of $\eta$ (i.e. the larger $a_1$ or $n$), the slower the convergence
and the sooner it deteriorates (i.e. the smaller the critical value of $M$ at which a rapid loss of accuracy occurs).
This issue is especially pronounced when increasing $N$,
which is compatible with a previous statement that the higher the order $j$ of $G_j$ (combined with higher resolution $N$),
the more drastically Fourier multipliers such as $C_j^{-|D|}$ can amplify round-off errors.
This result suggests that prescribing unnecessarily fine resolutions is not recommended.

\begin{figure}
\centering
\subfloat[]{\includegraphics[width=.5\linewidth]{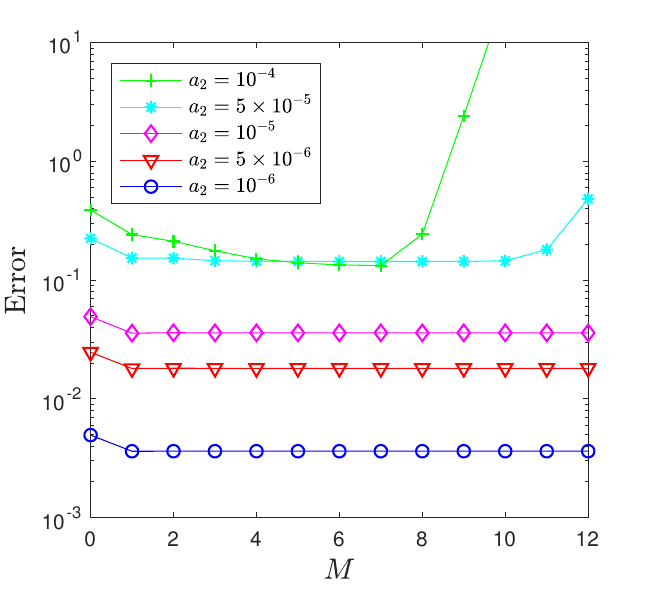}}
\hfill
\subfloat[]{\includegraphics[width=.5\linewidth]{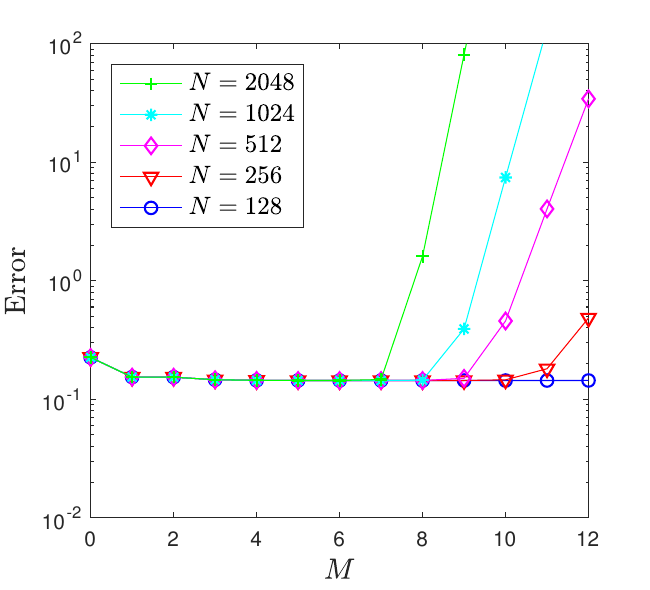}}
\caption{Relative $L^2$ error on the DNO vs. truncation order $M$ for $\eta$ given by \eqref{eta2}.
Graphs are shown for (a) varying amplitude $a_2$ (with $n = 2$, $N = 256$) 
and (b) varying resolution $N$ (with $a_2 = 5 \times 10^{-5}$, $n = 2$).}
\label{dno_k2_x4}
\end{figure}

\begin{figure}
\centering
\subfloat[]{\includegraphics[width=.5\linewidth]{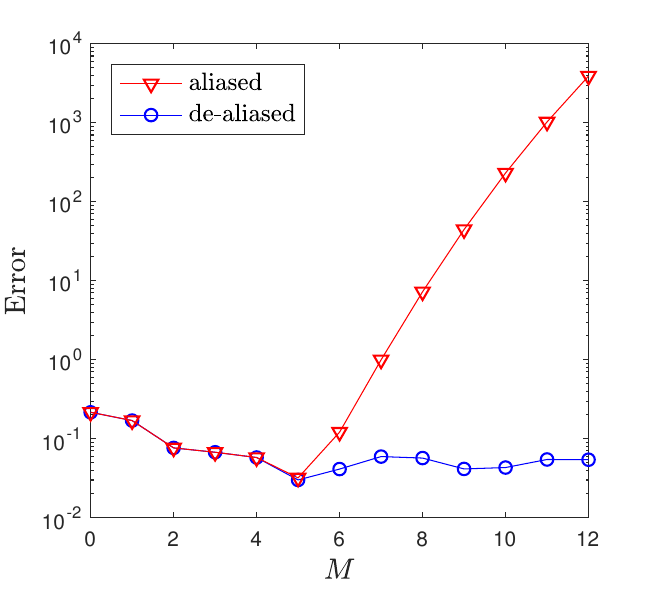}}
\hfill
\subfloat[]{\includegraphics[width=.5\linewidth]{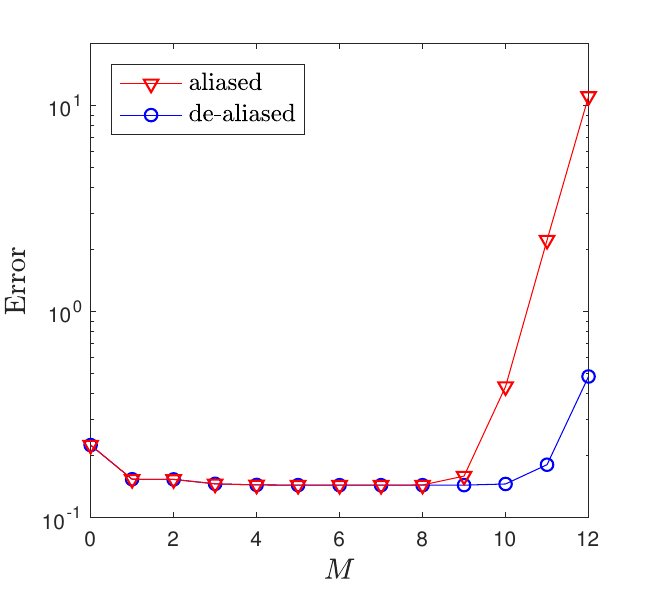}}
\caption{Relative $L^2$ error on the DNO vs. truncation order $M$ for $\eta$ given by 
(a) profile \eqref{eta1} (with $a_1 = 0.1$, $n = 10$) and (b) profile \eqref{eta2}
(with $a_2 = 5 \times 10^{-5}$, $n = 2$).
In each panel, the aliased and de-aliased results are compared.
In all these cases, the resolution is $N = 256$.}
\label{dno_N256_alias}
\end{figure}

From Fig. \ref{dno_k2_x4}, the same observations can be made on error plots for the surface profile \eqref{eta2}.
Because the bubble shape is oval-like in this case with $\eta$ being independent of $n$, we simply set $n = 2$ in \eqref{exact}
and skip any convergence test by varying $n$.
As expected, for this rougher boundary, the numerical issues appear to be more severe when increasing $a_2$ or $N$.
For reference, the maximum distortion associated with $a_2 = \{ 10^{-6}, \dots, 10^{-4} \}$ is 
$\| \eta \|_\infty = \{ 5 \times 10^{-3}, \dots, 5 \times 10^{-1} \}$ respectively, as depicted in Fig. \ref{dno_k2_x4}(a).
Finally, we illustrate in Fig. \ref{dno_N256_alias} the effectiveness of our de-aliasing procedure for both \eqref{eta1} and \eqref{eta2}
by comparing the corresponding $L^2$ errors with those obtained from aliased computations (without zero-padding).
Because aliasing typically arises when evaluating nonlinear terms and affects the high-wavenumber tail of the discretized spectrum,
not surprisingly it accentuates ill-conditioning effects at large values of $M$ and is more pronounced for higher surface steepnesses.
Representative cases are displayed in Fig. \ref{dno_N256_alias} for $(a_1, n) = (0.1, 10)$ 
and $(a_2, n) = (5 \times 10^{-5}, 2)$ with $N = 256$.

In summary, it may be inferred from these convergence tests that choosing a value around $M = 6$ (say, $4 \leq M \leq 8$)
would generally be a good compromise between accuracy and efficiency for the DNO approximation.
Similar values of $M$ have been successfully implemented in other physical contexts \cite{gn07,gp12,gp16,xg09}.

\subsection{Comparison with the Rayleigh--Plesset equation}

In this section, we further assess the performance of our boundary perturbation approach by comparing 
time-dependent simulations of \eqref{system}--\eqref{Lv} against predictions from the RP equation \eqref{RPeq}.
For this purpose, the second-order nonlinear ODE is transformed to a system of two first-order ODEs 
\begin{eqnarray} \label{RP1}
\frac{ds}{dt} & = & v \,, \\
\frac{dv}{dt} & = & \frac{1}{s \ln(R_m/s)} \left[v^2 \left( \frac{1}{2} - \ln \left( \frac{R_m}{s} \right) \right)
- \frac{\sigma}{\rho s} - \frac{p_\infty}{\rho} + \frac{1}{\rho} \left( p_\infty 
+ \frac{\sigma}{R} \right) \left( \frac{R}{s} \right)^{2\gamma} \right] \,,
\label{RP2}
\end{eqnarray} 
which is integrated numerically in time via the fourth-order Runge--Kutta scheme as applied to \eqref{system}--\eqref{Lv}.
Equations \eqref{RP1}--\eqref{RP2} thus lead to a direct numerical solver for this nonlinear problem, 
modulo the restriction to purely circular motions.
Accordingly, the present tests may be viewed as complementary to those from the previous section in the sense that
the focus now is on \eqref{G0p}--\eqref{Gjp} for $G^{(2)}$ as it is the relevant contribution to the DNO.
Note that Eqs. \eqref{RP1}--\eqref{RP2} are completely equivalent to their Hamiltonian counterparts \eqref{eq_q}--\eqref{eq_p}
but were preferred for our simulations because they directly yield the variable $s$ as an output.
For the sake of comparison, the same time step $\Delta t$ is used when solving 
\eqref{system}--\eqref{Lv} and \eqref{RP1}--\eqref{RP2}.
The same choice of non-dimensionalization is adopted for both systems of equations.

\begin{figure}
\centering
\subfloat[]{\includegraphics[width=.5\linewidth]{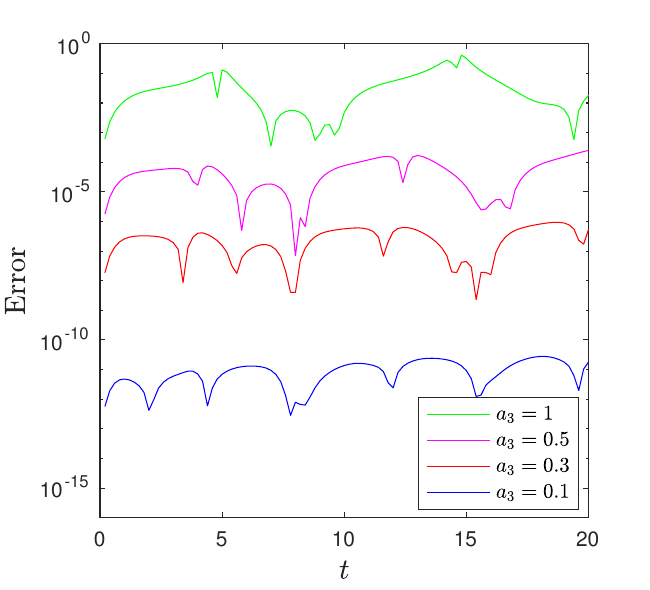}}
\hfill
\subfloat[]{\includegraphics[width=.5\linewidth]{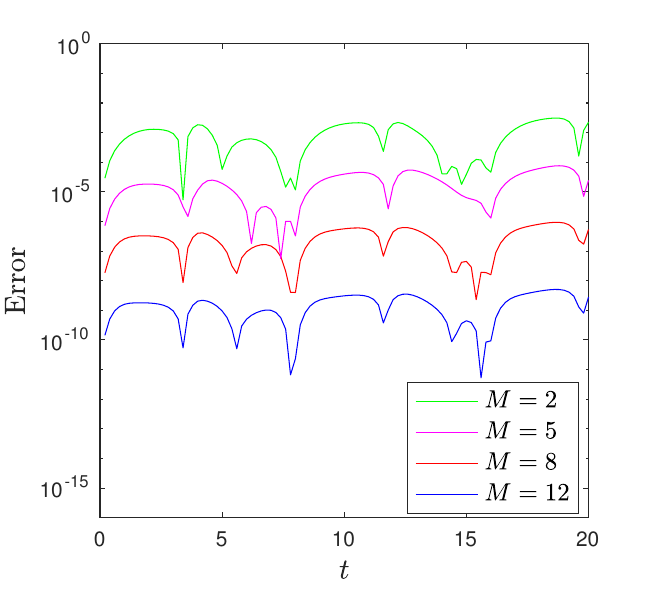}}
\caption{Relative error on bubble radius $s$ vs. time $t$ for far-field pressure $p_\infty = 1$ and initial conditions given by \eqref{init3}.
Graphs are shown for (a) varying initial amplitude $a_3$ (with $M = 8$, $N = 256$) 
and (b) varying truncation order $M$ (with $a_3 = 0.3$, $N = 256$).}
\label{RPcomp_N256}
\end{figure}

In the following illustrative experiments, we set $\gamma = 1.4$ under isentropic conditions \cite{d04,lh89}
and assume a uniform far-field pressure, say $p_\infty = 1$ in dimensionless units, for simplicity.
Again, the parameter $R_m = 1000 R$ is prescribed sufficiently large to avoid any possible logarithmic singularity in \eqref{RP2}.
Figure \ref{RPcomp_N256} shows the relative error
\begin{equation} \label{error2}
{\rm Error} = \left| \frac{R + \eta(0,t) - s(t)}{s(t)} \right| \,,
\end{equation}
between $\eta$ from \eqref{system}--\eqref{Lv} and $s$ from \eqref{RP1}--\eqref{RP2}, as a function of time $t$.
Due to angular invariance in this case, data on $\eta$ from \eqref{system}--\eqref{Lv} can be taken
at any arbitrary angle $\theta$, say $\theta = 0$, when evaluating \eqref{error2}.
For both systems of equations, the same initial conditions are specified at $t = 0$, namely
\begin{equation} \label{init3}
\big( \eta(\theta,0), \xi(\theta,0) \big) = (a_3, 0) \,, \quad \big( s(0), v(0) \big) = (R + a_3,0) \,,
\end{equation}
with $\Delta t = 0.01$ to integrate in time.
Errors \eqref{error2} are plotted in Fig. \ref{RPcomp_N256}(a) for varying $a_3$ (with $M = 8$, $N = 256$) 
and in Fig. \ref{RPcomp_N256}(b) for varying $M$ (with $a_3 = 0.3$, $N = 256$).
Convergence with respect to surface amplitude (here the initial amplitude $a_3$) or truncation order $M$ is confirmed by these results,
i.e. the lower $a_3$ or the larger $M$, the smaller the error at any time $t$.
This contrasts with the quick stagnation of errors when increasing $M$, 
as reported in Sec. \ref{Convergence} for $\theta$-dependent surface profiles.
Along the lines of our previous comment on this point, the better outcome here may be attributed to $G^{(1)}$ being inactive
and the absence of such Fourier multipliers as $C_j^{-|D|}$ from \eqref{Gjp},
thus making $G^{(2)}$ better conditioned for purely circular applications.
Even in the large-deflection regime $a_3 = 1$ (being on the order of $R$), the errors still remain acceptable (below $100$\%).
Because Eqs. \eqref{system}--\eqref{Lv} are tested against the RP system \eqref{RP1}--\eqref{RP2}
which is itself solved numerically (and thus is also affected by truncation or round-off errors, though to a lesser extent),
error fluctuations over time are expected as displayed in Fig. \ref{RPcomp_N256}.
Beside the low levels of these instantaneous errors, their overall stable behavior over time
attests further to the accuracy and effectiveness of our boundary perturbation approach.
The computations were not found to be particularly sensitive to $N$ in this regular $\theta$-independent geometry,
so convergence tests with varying $N$ are not shown here for convenience.

\begin{figure}
\centering
\subfloat[$a_3 = 0.3$]{\includegraphics[width=.5\linewidth]{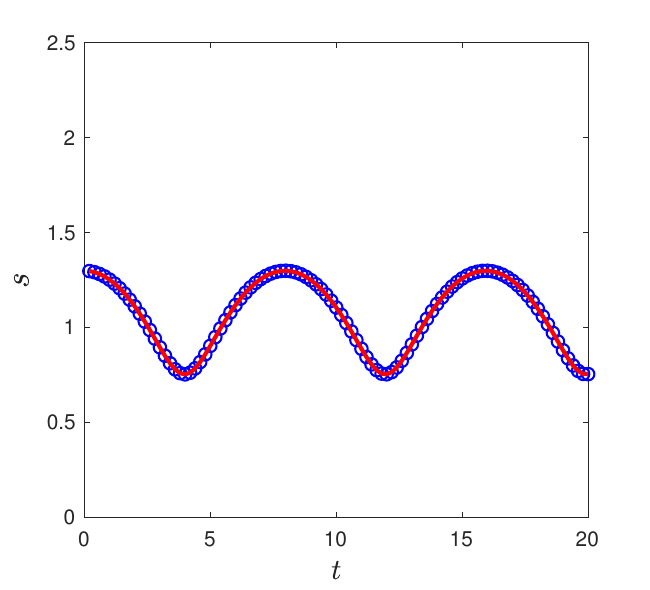}}
\hfill
\subfloat[$a_3 = 1$]{\includegraphics[width=.5\linewidth]{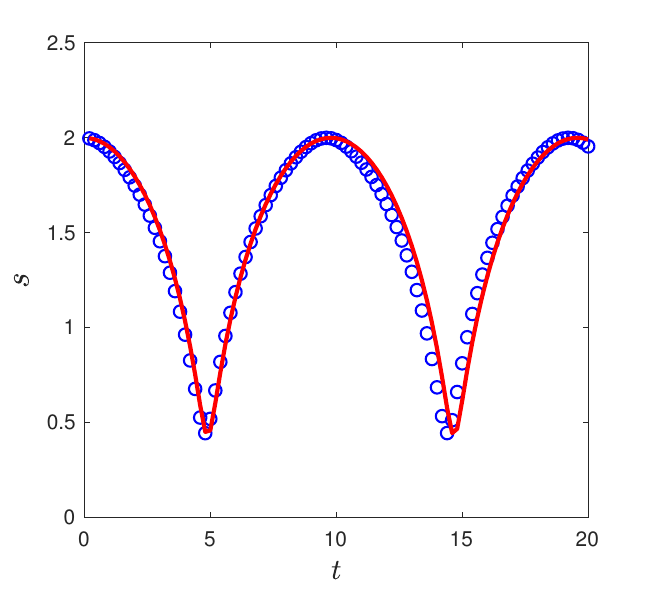}}
\caption{Bubble radius $s$ vs. time $t$ for far-field pressure $p_\infty = 1$ and initial conditions given by \eqref{init3}.
Graphs are shown for (a) $a_3 = 0.3$ and (b) $a_3 = 1.0$ (with $M = 8$, $N = 256$).
For each $a_3$, predictions from \eqref{system}--\eqref{Lv} (blue circles) and \eqref{RP1}--\eqref{RP2} (red line) are compared.}
\label{radcomp_M8_N256}
\end{figure}

Graphs of the bubble radii predicted by these two models during the time evolution are compared in Fig. \ref{radcomp_M8_N256} 
for $a_3 = 0.3$ and $1.0$ (with $M = 8$, $N = 256$).
The competition between $p_\infty$ and $p_B$ induces a cycle of compression-dilatation around the equilibrium state $r = R$,
with surface deflections of maximum amplitude $a_3$ (as determined by the initial amplitude) occurring at the dilatation peaks.
This cycle of compression-dilatation is asymmetric in the sense that the surface amplitude (relative to $r = R$) 
at the compression dips is slightly smaller than $a_3$.
Moreover, the dynamics during compression seems to be faster than that during dilatation,
as suggested by the steeper dips and smoother peaks along the curves of Fig. \ref{radcomp_M8_N256}.
A small discrepancy between these two estimated radii is discernible around $t = 14$ (near a compression dip) for $a_3 = 1.0$,
which is consistent with the larger error as revealed by Fig. \ref{RPcomp_N256}(a) in this case. 
The cyclic period is graphically inferred to be $\tau = 8.0$ and $9.8$ for $a_3 = 0.3$ and $1.0$ respectively,
which is comparable to the linear fundamental period $\tau_0 = 2\pi/\Omega_0 = 7.7$ of radial oscillations,
with $\Omega_0$ given by \eqref{RPdisp} for $\gamma = 1.4$ and $p_\infty = 1$ based on the RP equation \eqref{RPeq}.
The slight deviation is attributable to nonlinear effects, considering that $\tau$ is found to increase with $a_3$.
If the far-field pressure $p_\infty$ was amplified, the bubble dynamics would retain similar features 
but the cyclic period would be shorter, in accordance with the dependence of \eqref{RPdisp} on $p_\infty$.
We point out that filtering was not used in any of these simulations, for either \eqref{system}--\eqref{Lv} or \eqref{RP1}--\eqref{RP2}.

\begin{figure}
\centering
\includegraphics[width=.7\linewidth]{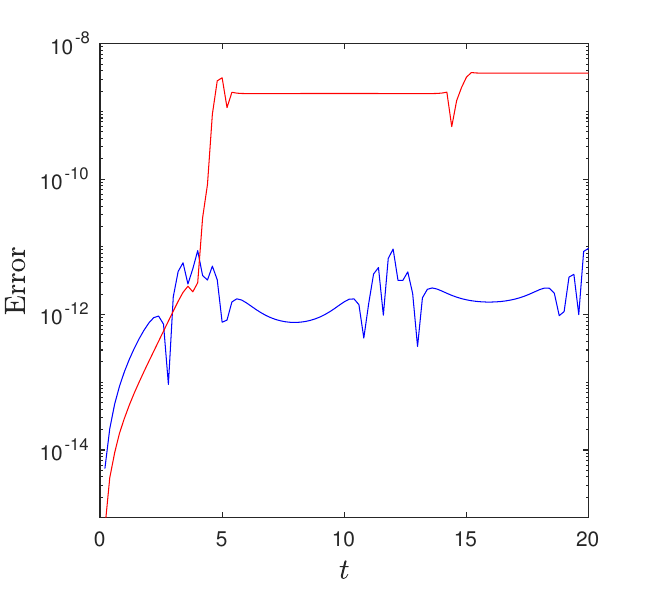}
\caption{Relative error on ${\cal H}$ vs. time $t$ for far-field pressure $p_\infty = 1$ and initial conditions given by \eqref{init3}.
Predictions from the RP model \eqref{RP1}--\eqref{RP2} are shown for $a_3 = 0.3$ (blue curve)
and $a_3 = 1$ (red curve) with $\Delta t = 0.01$.}
\label{RPener_dt001_M8_N256}
\end{figure}

The canonical conjugate variables $(\mathrm{q},\mathrm{p})$ for the RP equation \eqref{RPeq} 
can be reconstructed from the simulated variables $(s,v)$ according to \eqref{RPcoord},
so that the associated Hamiltonian \eqref{RPhamil} can be evaluated.
The conservation of this Hamiltonian ${\cal H}$ is confirmed by Fig. \ref{RPener_dt001_M8_N256}
which plots the time evolution of the error
\[
{\rm Error} = \left| \frac{{\cal H} - {\cal H}_0}{{\cal H}_0} \right| \,, 
\]
relative to the initial value ${\cal H}_0$, for $a_3 = 0.3$ and $1.0$.
Recall that this Hamiltonian structure is specific to the RP equation \eqref{RPeq}
and is not a direct consequence of the Hamiltonian property \eqref{canonical} exhibited by the full system.
In particular, ${\cal H}$ remains a conserved quantity under the action of $\Delta p$
(with uniform $p_\infty$) as given by \eqref{RPhamil},
while the original Hamiltonian formulation \eqref{canonical} with \eqref{ener2} does not extend to such a forced regime.

For computations in this purely circular setting, the RP model \eqref{RP1}--\eqref{RP2} is clearly a better option 
than \eqref{system}--\eqref{Lv}, owing to its simpler and more explicit form.
Nevertheless, this discussion helps validate \eqref{system}--\eqref{Lv} as a viable alternative to the RP equation,
highlighting the role of $G_j^{(2)}$ which can be readily incorporated into the DNO \eqref{G0t}--\eqref{Gjt}
for the general problem \eqref{laplace}--\eqref{farfield}, to produce a nonlinear model capable of describing a variety of bubble shapes.

\subsection{Simulation of nonlinear rotating waves}

Further examining the range of applicability of this boundary perturbation approach,
we now turn our attention to the simulation of nonlinear waves rotating steadily on the bubble surface.
Unlike the previous case, these shape distortions exhibit angular dependence according to \eqref{traveling}.
Equations \eqref{kine3}--\eqref{dyna3} constitute a boundary value problem with periodic boundary conditions in $\Theta \in [0,2\pi)$.
Via a pseudo-spectral method as outlined in Sec. \ref{Spectral},
$2 N$ discrete equations arise for $2 N$ unknowns $(\eta_j, \xi_j)$.
They are solved iteratively after specifying $c$ along with the initial guess \eqref{asymptotic}
based on second-order Stokes theory.
Through this initial guess, the parameter $n$ selects the wavenumber of the nonlinear wave.
The initial amplitude $a_0$ may be tuned depending on the choice of $c$ to help find the fixed-point solution.
This task is accomplished by using the Matlab routine \textit{fsolve},
which is essentially a quasi-Newton method with a numerical approximation of the Jacobian matrix.
We have successfully employed this Matlab routine in previous work \cite{g06,gp14,xg22} 
to obtain numerical predictions from nonlinear algebraic or differential equations.
Given a value of $n > 1$, we look here for a nonlinear wave rotating at speed $c > c_0$.
By analogy with the water wave problem \cite{f85}, an asymptotic behavior of the form
\[
c \simeq c_0 \left( 1 + \frac{1}{2} \varepsilon^2 \right) 
\simeq c_0 \left[ 1 + \frac{1}{2} \left( \frac{a_0 n}{R} \right)^2 \right] \,,
\]
may be anticipated according to Stokes theory up to third order
(recall that the second-order contribution $c_1 = 0$ as shown in Sec. \ref{Stokes}),
which offers the following estimate
\[
a_0 \simeq \frac{R}{n} \sqrt{2 \left( \frac{c}{c_0} - 1 \right)} \,,
\]
for $a_0$ in terms of $c$ and $n$, to be prescribed in the initial guess \eqref{asymptotic}.

\begin{figure}
\centering
\subfloat[$c = 1.64$]{\includegraphics[width=.45\linewidth]{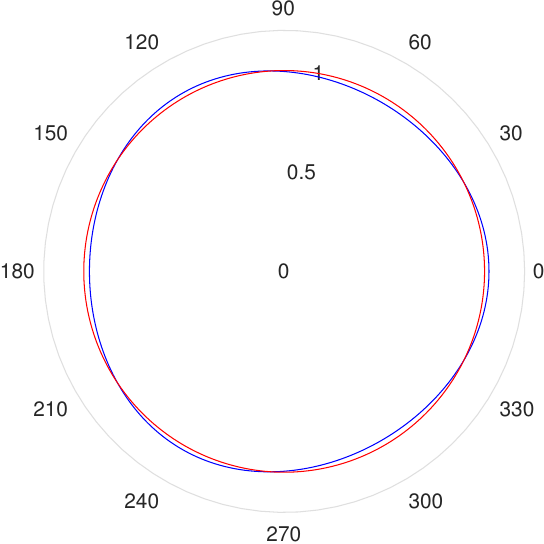}}
\hfill
\subfloat[$c = 1.70$]{\includegraphics[width=.45\linewidth]{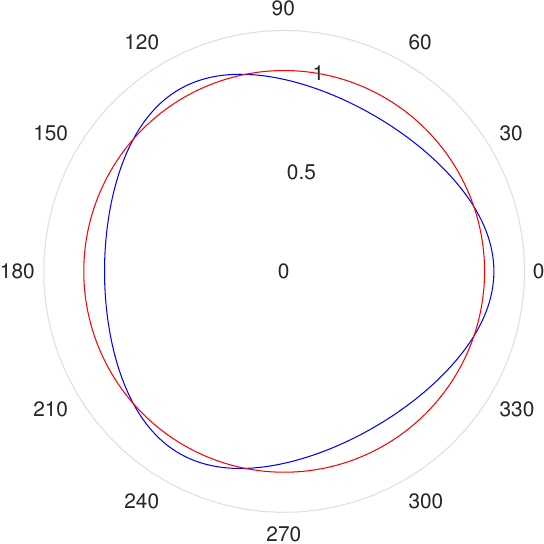}}
\hfill
\subfloat[$c = 1.75$]{\includegraphics[width=.45\linewidth]{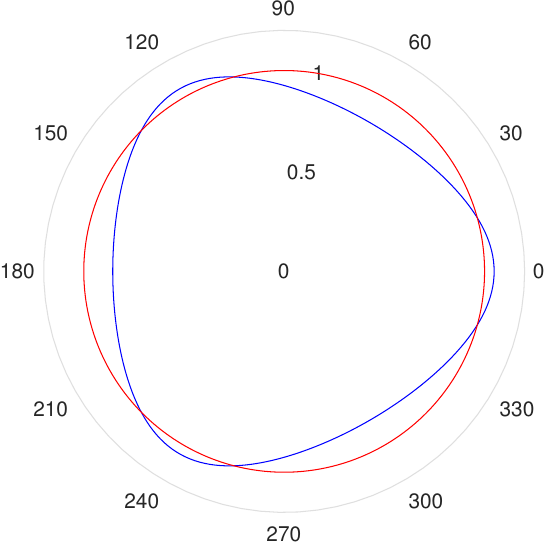}}
\hfill
\subfloat[$c = 1.86$]{\includegraphics[width=.45\linewidth]{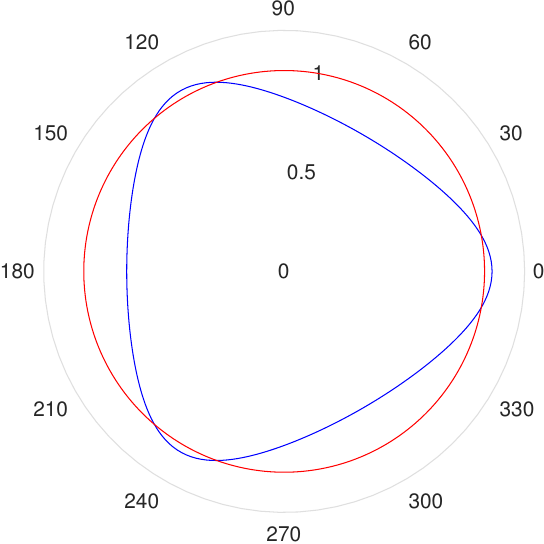}}
\caption{Bubble shape $s$ in polar coordinates $(r,\Theta)$ for steadily rotating waves with $n = 3$ and $N = 256$. 
The different panels correspond to angular speeds (a) $c = 1.64$, (b) $c = 1.70$, (c) $c = 1.75$, (d) $c = 1.86$.
As a reference, the red circle represents the unperturbed bubble of radius $R = 1$. Angles are indicated in degrees.}
\label{profile_trav_k3}
\end{figure}

Figure \ref{profile_trav_k3} depicts the bubble shapes produced by such rotating waves
for wavenumber $n = 3$ and angular speeds $c = \{ 1.64, 1.70, 1.75, 1.86 \}$.
In this case, the linear phase speed is $c_0 = 1.63$.
These nonlinear solutions were computed using $N = 256$ and $M = 4$ or $6$
for \eqref{series2} with \eqref{G0}--\eqref{Gj}.
Indeed, for wave speeds higher than $c = 1.70$, it was necessary to specify
a lower truncation order of the DNO ($M = 4$ rather than $M = 6$) in order to ensure convergence
of the fixed-point iterative scheme.
This issue is likely related to ill-conditioning of the DNO computation as discussed in Sec. \ref{Convergence}.
For high wave speeds, we also needed to include intermediate steps (with intermediate wave speeds)
as part of the iterative process going from the initial $c_0$ to the target $c$,
where the initial guess at each step is given by the converged solution from the previous step.
This strategy improves the convergence as compared to a single \textit{fsolve} search from $c_0$ to $c$.
We can discern in Fig. \ref{profile_trav_k3} that the higher $c$, the larger the wave amplitude and steepness
(with steeper crests vs. smoother troughs).
More quantitatively, the amplitude of steadily rotating waves can be evaluated as
$a = |\eta_{\max} - \eta_{\min}|/2$.
For $c = \{ 1.64, 1.70, 1.75, 1.86 \}$, we find $a = \{ 0.02, 0.07, 0.09, 0.12 \}$ respectively,
hence wave steepness $\varepsilon = (a \, n)/R = \{ 0.07, 0.22, 0.28, 0.37 \}$
which is a measure of the nonlinearity strength.
These features are clearly revealed by Fig. \ref{profile2dc_trav}(a)
where the bubble's surface deformations are plotted versus $\Theta$ in a rectangular format.
These graphs also indicate that the wave steepening is accompanied by 
a decrease of mean surface level $Q$, hence a decrease of bubble volume $V$.

\begin{figure}
\centering
\subfloat[$c = 1.95$]{\includegraphics[width=.45\linewidth]{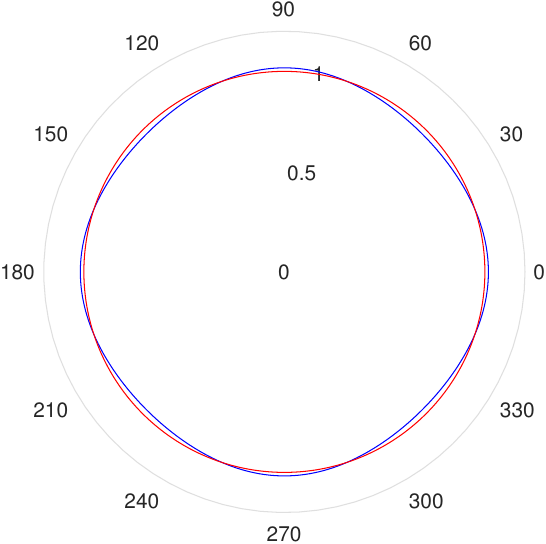}}
\hfill
\subfloat[$c = 2.00$]{\includegraphics[width=.45\linewidth]{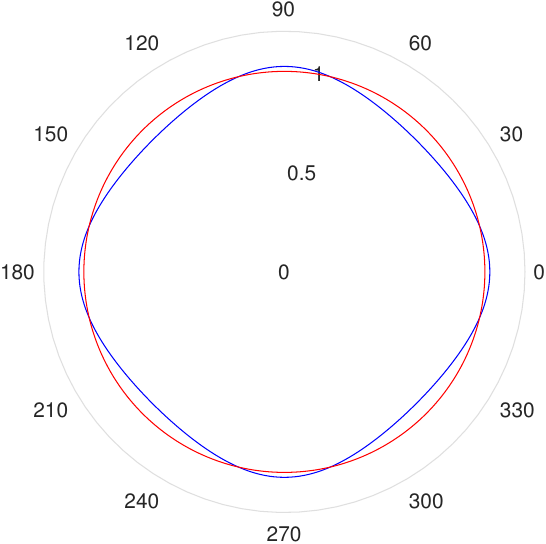}}
\hfill
\subfloat[$c = 2.10$]{\includegraphics[width=.45\linewidth]{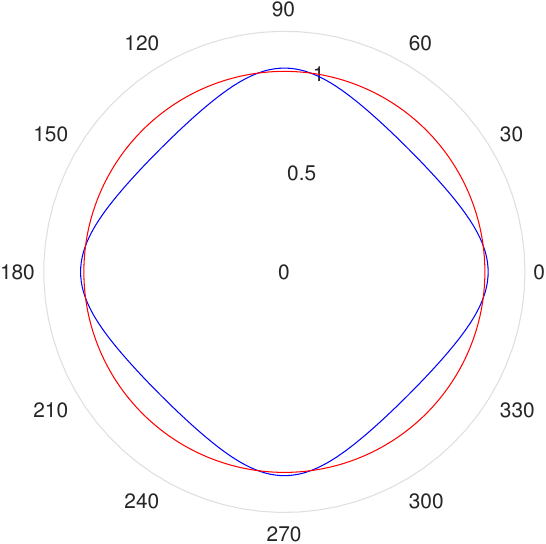}}
\hfill
\subfloat[$c = 2.29$]{\includegraphics[width=.45\linewidth]{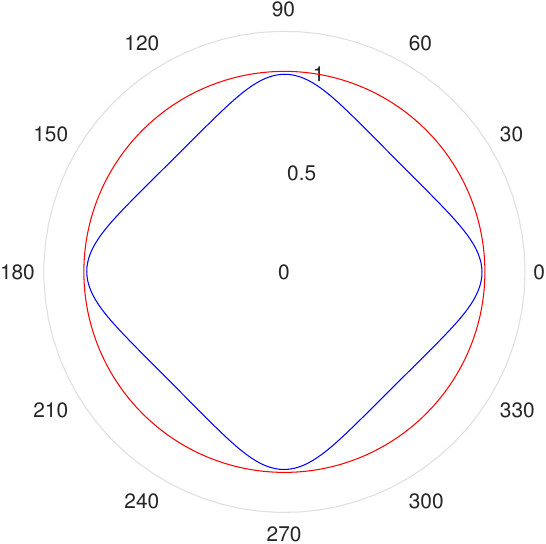}}
\caption{Bubble shape $s$ in polar coordinates $(r,\Theta)$ for steadily rotating waves with $n = 4$ and $N = 256$. 
The different panels correspond to angular speeds (a) $c = 1.95$, (b) $c = 2.00$, (c) $c = 2.10$, (d) $c = 2.29$.
As a reference, the red circle represents the unperturbed bubble of radius $R = 1$. Angles are indicated in degrees.}
\label{profile_trav_k4}
\end{figure}

\begin{figure}
\centering
\subfloat[$n = 3$]{\includegraphics[width=.5\linewidth]{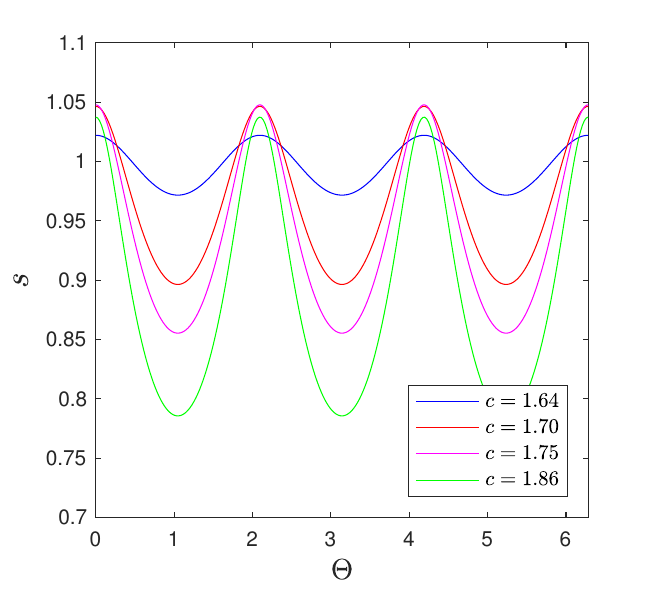}}
\hfill
\subfloat[$n = 4$]{\includegraphics[width=.5\linewidth]{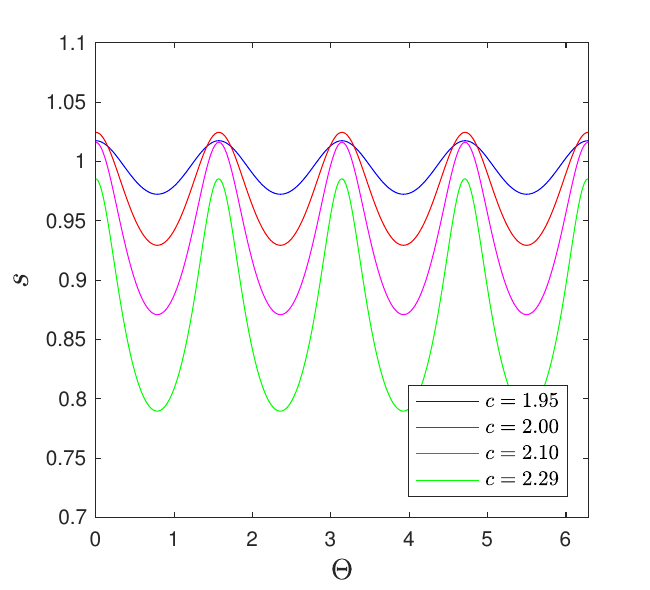}}
\caption{Bubble deformation $s$ vs. angle $\Theta$ for steadily rotating waves with (a) $n = 3$ and (b) $n = 4$.
In each panel, graphs are shown for varying speed $c$ with $N = 256$.}
\label{profile2dc_trav}
\end{figure}

Similar observations can be made when inspecting computations of mode-4 ($n = 4$) rotating waves
as portrayed in Figs. \ref{profile_trav_k4} and \ref{profile2dc_trav}(b).
For $c = \{ 1.95, 2.00, 2.10, 2.29 \} > c_0 = 1.93$ in this setting,
we find $a = \{ 0.02, 0.04, 0.07, 0.09 \}$ and $\varepsilon = \{ 0.09, 0.19, 0.29, 0.39 \}$ respectively.
While the bubble under $n = 3$ wave excitation tends to exhibit a triangular-like shape as $c$ increases,
the limiting geometry for $n = 4$ is more square-like.
In either case, we were not able to achieve convergence for wave speeds higher than those considered
in Fig. \ref{profile2dc_trav}.

\begin{figure}
\centering
\subfloat[$n = 3$]{\includegraphics[width=.5\linewidth]{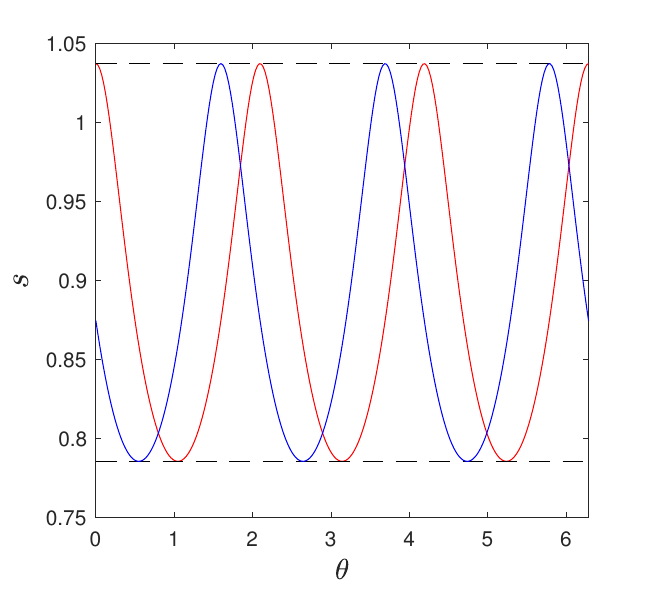}}
\hfill
\subfloat[$n = 4$]{\includegraphics[width=.5\linewidth]{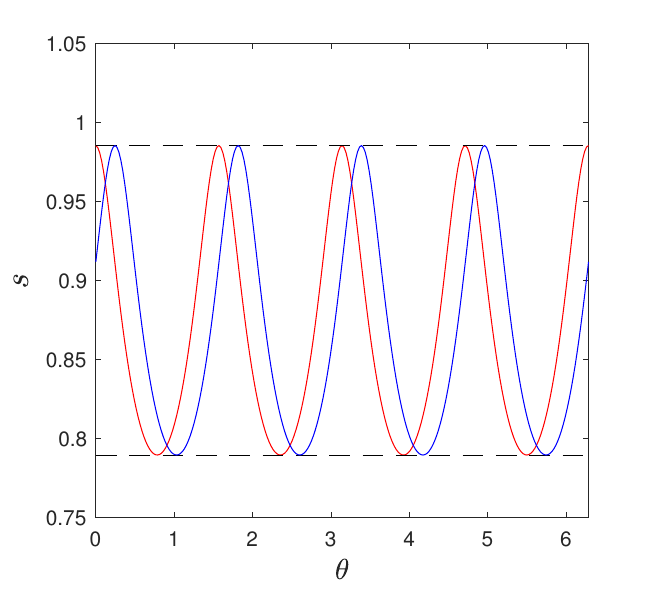}}
\caption{Bubble deformation $s$ vs. angle $\theta$ at $t = 0$ (red line) and $t = 20$ (blue line) 
during the propagation of steadily rotating waves for (a) $c = 1.86$, $n = 3$ 
and (b) $c = 2.29$, $n = 4$ (with $M = 4$, $N = 256$).}
\label{wave_M4_N256}
\end{figure}

We further check the accuracy of our fixed-point iterative algorithm for \eqref{kine3}--\eqref{dyna3}
by prescribing its converged solutions as initial conditions for \eqref{system}--\eqref{Lv}
and by solving these evolution equations via the time integration method described in Sec. \ref{Integration}.
Under the same conditions as in Sec. \ref{Stokes}, their implementation is restricted to \eqref{G0}--\eqref{Gj}
for $G(\eta)$ and to \eqref{Young} for $\Delta p$.
The steady character of these rotating waves is confirmed by Fig. \ref{wave_M4_N256}
where snapshots of the bubble deformation are shown at $t = 0$, $20$ in the two highly nonlinear regimes
$c = 1.86$ ($n = 3$) and $c = 2.29$ ($n = 4$).
For both solutions, the surface profiles at $t = 0$ and $t = 20$ look indistinguishable,
modulo a phase shift due to the uniform wave propagation.
While these snapshots happen to be located close together in either plot,
it should be kept in mind that the wave has actually rotated multiple times 
across the computational domain over the span of the simulation, because of the periodic boundary conditions.
The associated numerical parameters are $\Delta t = 0.001$, $M = 4$ and $N = 256$.
For such high wave speeds, filtering was required to run the time-dependent computations as explained earlier
(see Sec. \ref{Filtering}).

\begin{figure}
\centering
\subfloat[$n = 3$]{\includegraphics[width=.5\linewidth]{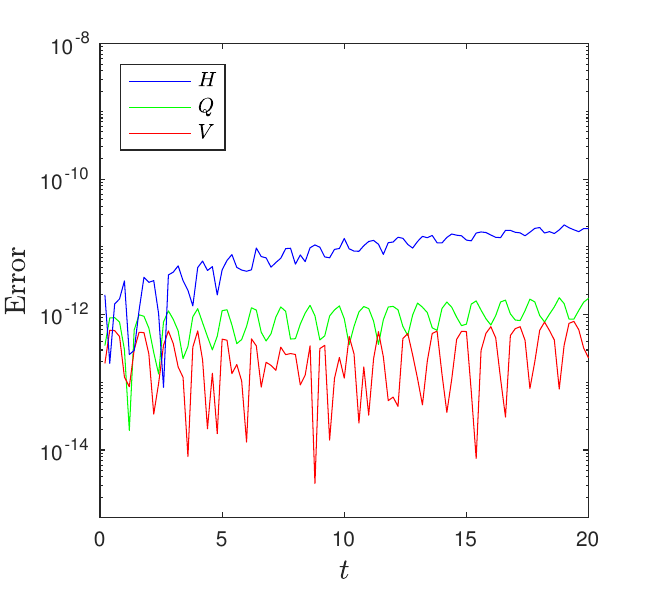}}
\hfill
\subfloat[$n = 4$]{\includegraphics[width=.5\linewidth]{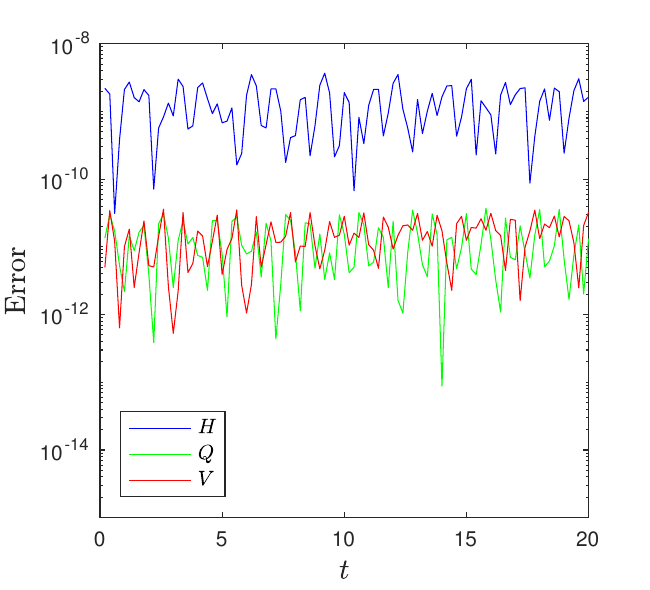}}
\caption{Relative errors on $H$, $Q$, $V$ vs. time $t$ during the propagation of steadily rotating waves 
for (a) $c = 1.86$, $n = 3$ and (b) $c = 2.29$, $n = 4$ (with $M = 4$, $N = 256$).}
\label{ener_M4_N256}
\end{figure}

A more quantitative assessment of these rotating wave solutions is provided in Fig. \ref{ener_M4_N256} 
which plots the relative errors
\[
{\rm Error} = \left| \frac{H - H_0}{H_0} \right| \,, \quad 
\left| \frac{Q - Q_0}{Q_0} \right| \,, \quad \left| \frac{V - V_0}{V_0} \right| \,, 
\]
on energy $H$, mean surface level $Q$ and volume $V$ versus time $t$ for
$c = 1.86$ ($n = 3$) and $c = 2.29$ ($n = 4$) as presented by Fig. \ref{wave_M4_N256}.
The quantities $H_0$, $Q_0$, $V_0$ denote the corresponding initial values at $t = 0$.
The trapezoidal rule was employed to evaluate integrals in the definitions 
\eqref{level}, \eqref{mass2}, \eqref{ener2} of these invariants.
Overall, the minuscule values of these errors (in particular near machine precision for $Q$ and $V$)
as well as their stable evolution over time confirms that $H$, $Q$, $V$ are very well conserved numerically
and that steadily rotating waves are simulated accurately.
The computation of $H$ being more prone to inaccuracy than $Q$ or $V$ is likely due to
its more complicated expression \eqref{ener2} which directly involves the DNO approximation.
Moreover, the slightly larger errors for $c = 2.29$ ($n = 4$, $\varepsilon = 0.39$) as compared to
$c = 1.86$ ($n = 3$, $\varepsilon = 0.37$) are expected considering that
our boundary perturbation approach is better suited for milder bubble deformations.

\subsection{Simulation of nonlinear standing waves}

Another class of interesting nonlinear solutions is that of time-periodic standing waves.
Such solutions have been much investigated in the three-dimensional axisymmetric configuration,
especially by means of analytical perturbation calculations \cite{tb83}.
Their direct numerical computation is particularly challenging in the nonlinear regime,
which may be tackled by solving the associated constrained problem via an adjoint method,
and thus is envisioned for future work.
Alternatively, a standing wave can be produced by superimposing two identical constant-amplitude waves
that rotate in opposite directions.

Exploiting this idea here, we numerically solve \eqref{system}--\eqref{Lv} in time together with \eqref{Young} 
from initial conditions given by
\[
\eta(\theta,0) = \eta_1(\theta) + \eta_2(\theta) \,, \quad
\xi(\theta,0) = \xi_1(\theta) - \xi_2(\theta) \,,
\]
at $t = 0$, where both pairs $(\eta_1,\xi_1)$ and $(\eta_2,\xi_2)$ denote steadily rotating waves
with the same speed $c > c_0$ and wavenumber $n > 1$, as discussed in the previous section, 
but with possibly a phase shift between them.
The minus sign in $\xi(\theta,0)$ above is to ensure that these two initial waves rotate in opposite directions.
Their interaction will create a standing (non-rotating) wave pattern that exhibits the same wavenumber $n$
as the original rotating waves but oscillates in time at a specific frequency $\omega$.
Because of this superposition, the resulting standing wave reaches an amplitude equal to
the sum of the initial amplitudes, which makes it a more nonlinear solution than each individual rotating wave.

\begin{figure}
\centering
\subfloat[$t = 0.00$]{\includegraphics[width=.3\linewidth]{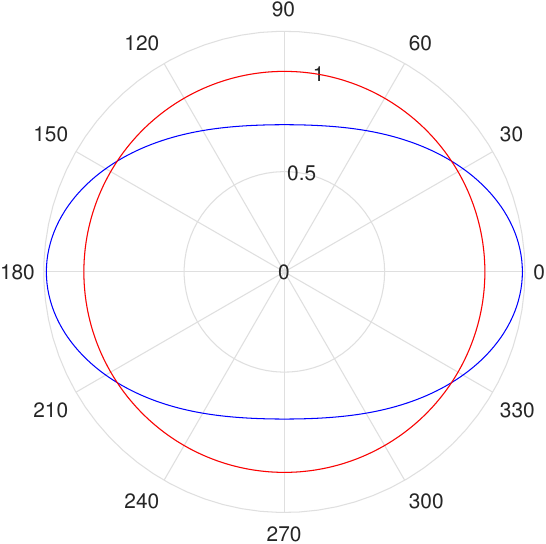}}
\hfill
\subfloat[$t = 0.34$]{\includegraphics[width=.3\linewidth]{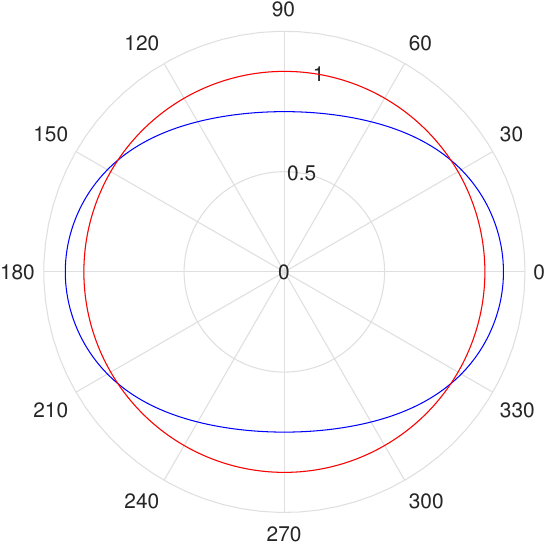}}
\hfill
\subfloat[$t = 0.50$]{\includegraphics[width=.3\linewidth]{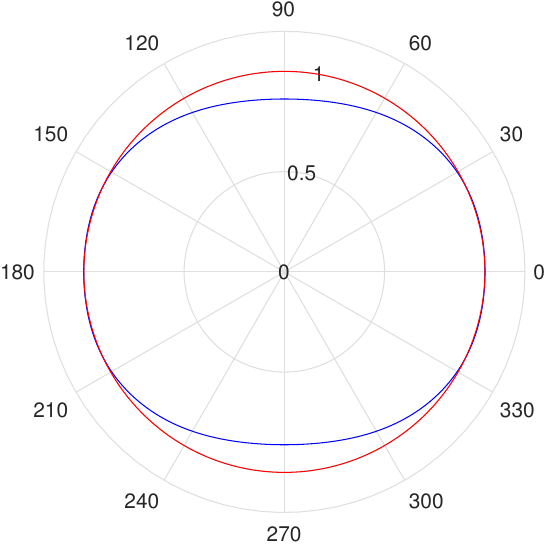}}
\hfill
\subfloat[$t = 0.62$]{\includegraphics[width=.3\linewidth]{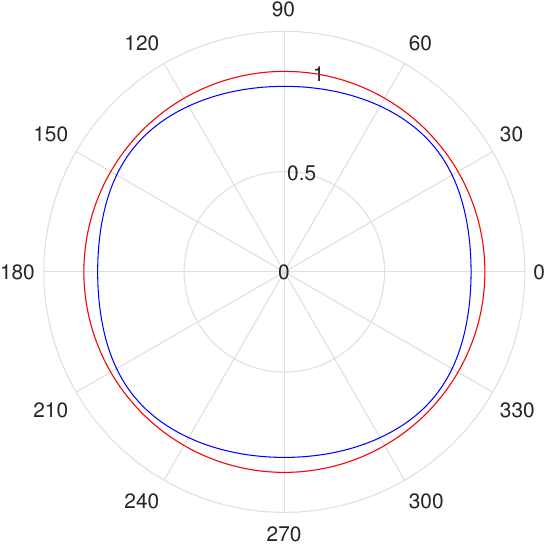}}
\hfill
\subfloat[$t = 0.76$]{\includegraphics[width=.3\linewidth]{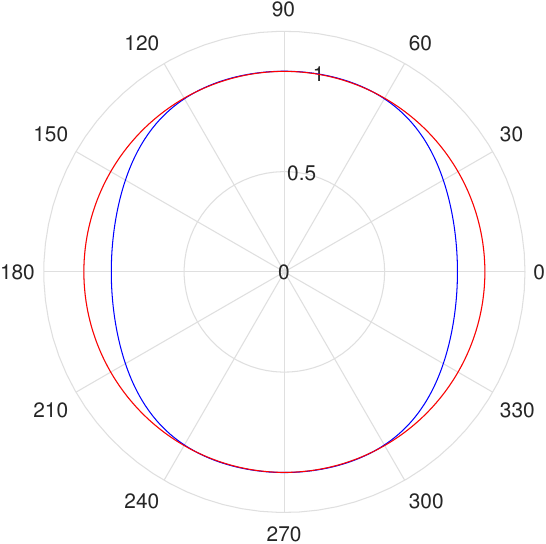}}
\hfill
\subfloat[$t = 0.94$]{\includegraphics[width=.3\linewidth]{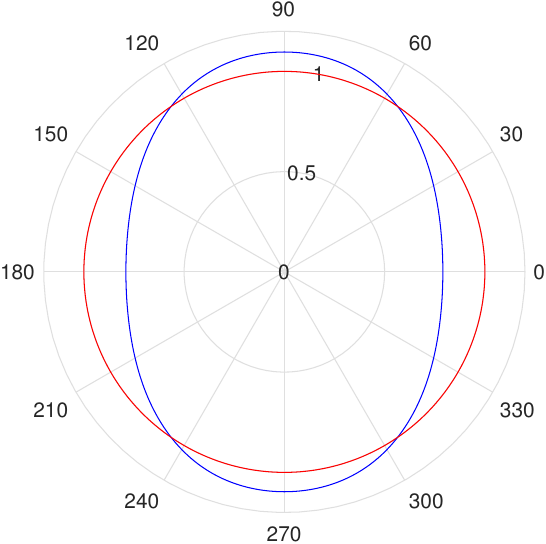}}
\hfill
\subfloat[$t = 1.26$]{\includegraphics[width=.3\linewidth]{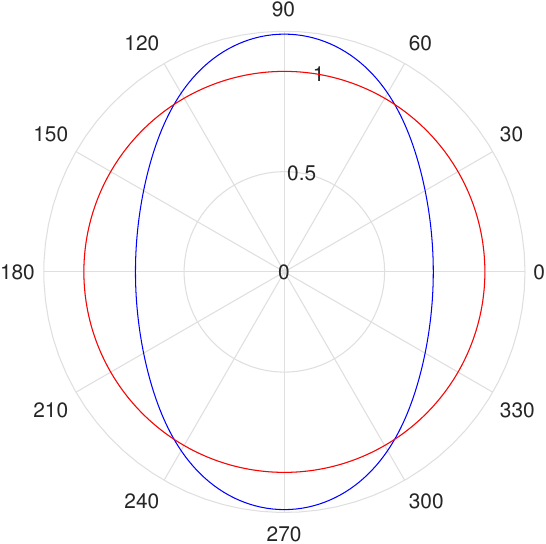}}
\hfill
\subfloat[$t = 2.52$]{\includegraphics[width=.3\linewidth]{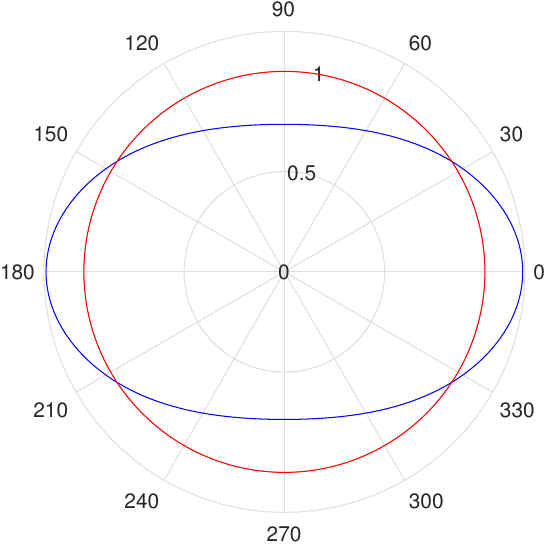}}
\hfill
\subfloat[$t = 3.78$]{\includegraphics[width=.3\linewidth]{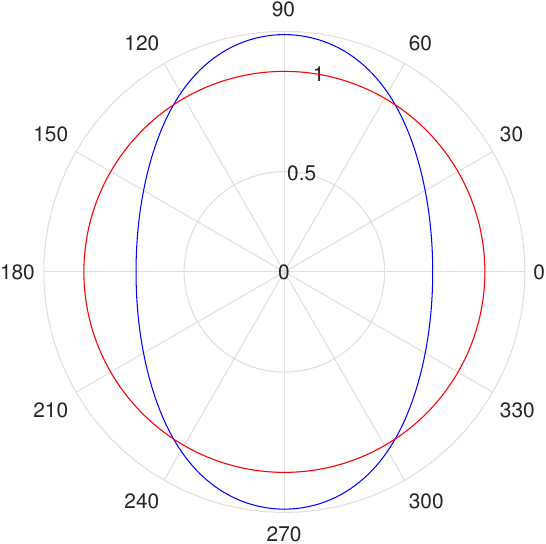}}
\caption{Bubble shape $s$ in polar coordinates $(r,\theta)$ for a time-periodic standing wave due to
the superposition of two steadily rotating waves with $c = 1.26$ and $n = 2$ ($M = 4$, $N = 256$). 
The different panels correspond to times (a) $t = 0.00$, (b) $t = 0.34$, (c) $t = 0.50$, (d) $t = 0.62$,
(e) $t = 0.76$, (f) $t = 0.94$, (g) $t = 1.26$, (h) $t = 2.52$, (i) $t = 3.78$.
As a reference, the red circle represents the unperturbed bubble of radius $R = 1$. Angles are indicated in degrees.}
\label{profile_stan_k2}
\end{figure}

Figure \ref{profile_stan_k2} illustrates this process with two initial rotating waves corresponding to $c = 1.26$ and $n = 2$.
They are initially positioned in phase to interfere with each other in such a way that oblate deformations
associated with the mode-2 standing wave occur along the axis $\theta = 0$ 
while prolate deformations occur along the axis $\theta = \pi/2$.
The bubble shapes at various instants are displayed in Fig. \ref{profile_stan_k2}
during the first cycle of oblate-prolate oscillation.
This cycle indeed repeats itself over time in our simulations.
Note the substantial surface deflection relative to the unit circle at the times of maximum oblate ($t = 0.00$, $2.52$)
and prolate ($t = 1.26$, $3.78$) oscillations.

\begin{figure}
\centering
\subfloat{\includegraphics[width=.7\linewidth]{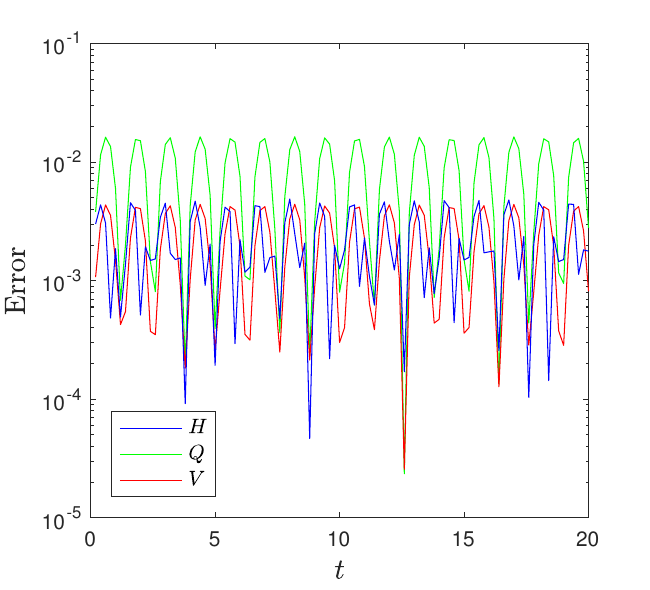}}
\caption{Relative errors on $H$, $Q$, $V$ vs. time $t$ during the evolution of a time-periodic standing wave
due to the superposition of two steadily rotating waves with $c = 1.26$ and $n = 2$ ($M = 4$, $N = 256$).}
\label{enerstand_k2_M4_N256}
\end{figure}

The favorable results on conservation of $H$, $Q$, $V$ as demonstrated by Fig. \ref{enerstand_k2_M4_N256}
attest to the computation accuracy in this standing wave case as well.
The idea that we synthetically superimpose two rotating waves at $t = 0$
(instead of prescribing an actual initial condition for the standing wave)
together with the fact that the resulting solution is highly nonlinear as pointed out earlier
may explain why we see higher error levels here than say, in Fig. \ref{ener_M4_N256} for a single rotating wave.

\begin{figure}
\centering
\subfloat{\includegraphics[width=.7\linewidth]{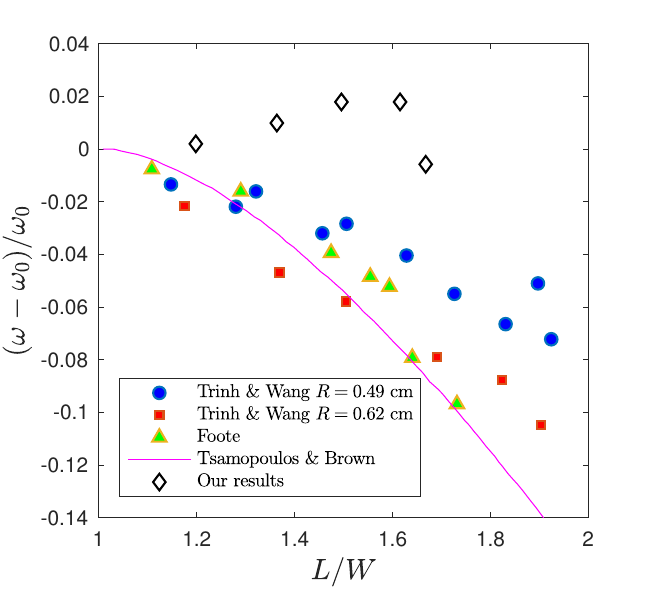}}
\caption{Frequency shift $(\omega - \omega_0)/\omega_0$ vs. aspect ratio $L/W$ at maximum deformation
for standing waves corresponding to $n = 2$ and $c = \{ 1.23, 1.24, 1.25, 1.26, 1.27 \}$ ($M = 4$, $N = 256$).
Our numerical results are compared to experimental data for drops by Trinh and Wang \cite{tw82},
numerical estimates for drops by Foote \cite{f73} and asymptotic predictions for bubbles by Tsamopoulos and Brown \cite{tb83}.}
\label{data_comp}
\end{figure}

Finally, we make an attempt to compare our numerical results in this setting with previous data from the literature.
Following Tsamopoulos and Brown \cite{tb83}, we plot in Fig. \ref{data_comp} estimates of the frequency shift
$(\omega - \omega_0)/\omega_0$ relative to $\omega_0$, the linear angular frequency \eqref{dispersion} with $n = 2$, 
versus the aspect ratio $L/W$ of the bubble's major axis $L$ to its minor axis $W$ at maximum prolate deformation.
We calculate $L/W$ from our simulations by tracking the bubble radius $s$ whereas, for the frequency $\omega$,
we first estimate the period $T$ of a prolate oscillation and then evaluate $\omega = 2\pi/T$.
In this two-dimensional context, the oblate shape dynamics turn out to exactly mirror the prolate shape dynamics,
exhibiting an identical cycle of oscillation with the same period $T$ and aspect ratio $L/W$.
Figure \ref{data_comp} shows a comparison with laboratory measurements by Trinh and Wang \cite{tw82}
for almost neutrally buoyant drops of silicon oil and carbon tetrachloride suspended in distilled water.
Data sets for two different drop radii $R = 0.49$ cm and $0.62$ cm are reported.
Numerical estimates by Foote \cite{f73} for drops as well as asymptotic predictions by Tsamopoulos and Brown \cite{tb83}
for bubbles are also included in Fig. \ref{data_comp}.
All these data were extracted from Fig. 4 in \cite{tb83}.
Note that this assessment is only meant to be qualitative because, on one hand, we compare two-dimensional simulations
with three-dimensional results for which quantitative differences are expected. 
In particular, the numerical or asymptotic results of Foote \cite{f73} and Tsamopoulos and Brown \cite{tb83}
assume a three-dimensional spherical geometry that is invariant under azimuthal rotation.
Secondly, our numerical estimates are based on inviscid potential-flow theory 
while laboratory measurements are subject to viscous effects.
Furthermore, drops and bubbles may reveal different behaviors depending on the situation,
and it should also be kept in mind that our computations were not set up under the same conditions
as in the laboratory experiments (regarding e.g. the wave excitation mechanism).

Our own data in Fig. \ref{data_comp} correspond to $c = \{ 1.23, 1.24, 1.25, 1.26, 1.27 \}$ representing standing waves
of amplitude $a = \{ 0.08, 0.15, 0.19, 0.22, 0.25 \}$ at maximum deformation.
The respective maximum wave steepnesses are $\varepsilon = \{ 0.17, 0.30, 0.38, 0.45, 0.50 \}$ for $n = 2$,
which is indicative of the highly nonlinear character of these computed solutions.
To this aim, we specified $M = 4$, $N = 256$ with a time step as small as
$\Delta t = 0.0002$ and $0.00001$ for $c = 1.26$ and $1.27$ respectively.
Despite this fine resolution in time, filtering was still required to stabilize the simulation of such large-amplitude waves
during their oscillatory evolution.

It is first remarked that our $L/W$ estimates fall within the range of values observed by these previous authors.
In our situation, the higher $c$, the larger $a$ (or $\varepsilon$) and so is the aspect ratio $L/W$ of standing waves.
We were not able to obtain two-dimensional solutions for $L/W$ values as high as those reported in three-dimensional configurations.
However, a striking difference from these previous studies is that we find a (positive) frequency upshift 
$(\omega - \omega_0)/\omega_0$ relative to $\omega_0$, rather than a (negative) frequency downshift.
Note that the negative ordinate for our last data point (the one farthest to the right associated with $c = 1.27$) is likely a numerical artifact.
Although our simulation managed to be completed in this highly nonlinear case (by means of filtering together with a small time step),
producing an acceptable solution overall, it was particularly subject to errors to the point that spurious noise
was detectable near the wave crests and troughs.

The qualitative agreement between the asymptotic predictions of Tsamopoulos and Brown \cite{tb83}
and the other data sets suggests that the discrepancy with our numerical results may be attributed to the two different geometries.
An argument in favor of this explanation stems from the different expression of the linear angular frequency
\begin{equation} \label{disp3D}
\omega_0^2 = \frac{\sigma}{\rho R^3} (n + 2) (n^2 - 1) \,, \quad n > 1 \,,
\end{equation}
for three-dimensional axisymmetric bubbles \cite{tb83}.
Clearly, $\omega_0$ from \eqref{disp3D} is larger than that from \eqref{dispersion} for any $n > 1$,
so the frequency shift $(\omega - \omega_0)/\omega_0$ may be more prone to being negative
in three dimensions than in two dimensions.
Another argument is related to the fact that three-dimensional drops or bubbles undergoing mode-2 oscillations
have been observed to spend a longer part of each period in a prolate form than in an oblate one,
which may be due to the vertical direction ($\theta = \pi/2$) being an axis of symmetry in the axisymmetric case \cite{tb83}.
By contrast, our two-dimensional problem is isotropic, giving rise to identical oblate and prolate shape dynamics as mentioned earlier.
Accordingly, the typical period $T$ of a nonlinear mode-2 standing wave in three dimensions may be longer than the linear period,
or equivalently the associated frequency $\omega$ may be shorter than $\omega_0$, 
hence a negative frequency shift $(\omega - \omega_0)/\omega_0$.
Finally, perhaps a simpler (more intuitive) argument is that, because a bubble volume is less in two dimensions
than in three dimensions for a given radius $R$, it thus takes less effort to excite the bubble surface at a certain mode $n$.
As a consequence, the excitation frequency $\omega$ should be larger in two dimensions than in three dimensions,
which promotes a frequency upshift in the former setting versus a frequency downshift in the latter setting.
Overall, this geometrical explanation for the different dynamics depending on the spatial dimension
is consistent with e.g. recent laboratory measurements by Duplat \cite{d19} who noticed that,
during the late stages of a circular bubble's collapse, its radius vanishes like $s(t) \sim (t_0 - t)^{1/2}$
as $t \to t_0$, which contrasts with the faster collapse like $s(t) \sim (t_0 - t)^{2/5}$ for a spherical bubble.

\section{Conclusions}

We consider the two-dimensional problem of free or forced deformations of a gas bubble immersed in a liquid of infinite extent.
This configuration has drawn much less attention than the three-dimensional axisymmetric geometry,
especially in applications to bubble acoustics.
Based on nonlinear theory for potential flow in the presence of a moving boundary under surface tension without gravity,
and adopting the Hamiltonian formulation proposed by Benjamin \cite{b87},
we introduce the DNO to clarify the reduction in terms of surface variables alone and to verify the canonical symplectic structure.
Benjamin \cite{b87} did not specifically treat this two-dimensional case in polar coordinates.

We propose a Taylor series representation of the DNO about a quiescent circular state of the bubble.
A recursion formula is devised to evaluate this Taylor series up to an arbitrary order of nonlinearity,
with each term having two distinct components to describe shape distortions and radial pulsations as well as their coupling.
Each of these components is given by a sum of concatenations of Fourier multipliers with powers of the surface displacement.

In this theoretical framework, various analytical approximations of the problem are examined.
We obtain a Stokes wave solution that is accurate up to second order in wave steepness 
for steadily rotating shape oscillations without volume change.
Assuming circular symmetry, we also derive an inviscid RP model for the time evolution of the bubble radius 
under the excitation of a far-field pressure.
Its independent Hamiltonian structure is established in the case of uniform pressure forcing.

Exploiting the lower-dimensional form of the full governing equations together with the Taylor expansion of the DNO,
we develop a direct numerical solver where each term in the DNO series is computed efficiently and accurately by
a pseudo-spectral method with the FFT.
Via this Fourier decomposition, an arbitrary number of shape modes can be specified to contribute to nonlinear wave interactions,
and the linear terms can be solved exactly by the integrating factor technique.
This approach is thus well suited for time-dependent simulations in a forced or freely evolving regime.

We show extensive numerical tests on the convergence of the DNO as a function of the truncation order $M$
for varying bubble shapes, surface steepnesses and spatial resolutions.
The significance of aliasing errors when computing nonlinear products is revealed 
and a zero-padding technique is implemented to remove them.
While the convergence errors remain small overall, they are not found to decrease exponentially with increasing $M$,
in contrast to what might be expected from the analyticity property of the DNO.
Rather, the convergence quickly stagnates past $M \simeq 2$ and even deteriorates past a critical value of $M$
for severe shape distortions or fine spatial resolutions.
This behavior is consistent with the ill-conditioning of the DNO series,
which has been documented in other contexts \cite{gp16,nr01,xg09}.

Despite these numerical issues associated with the DNO, computations of bubble deformations
of moderately large amplitude or steepness can be performed with satisfactory accuracy.
We apply this algorithm to simulating cycles of compression-dilatation for a purely circular bubble 
driven by a uniform pressure field.
We test these results against predictions by the RP model, thus providing mutual validation for these two proposed approaches.
Nonlinear shape oscillations of a non-circular bubble in the unforced regime without volume change are also investigated.
We compute steadily rotating waves and time-periodic standing waves for a few first modes of the bubble surface.
Focusing on mode-2 standing waves, we examine the shift in oscillation frequency (relative to the linear value)
as a function of the bubble's maximum aspect ratio, 
and we compare our numerical estimates to laboratory measurements and other theoretical predictions.
Our two-dimensional results show a frequency upshift that accentuates with increasing aspect ratio,
while previous studies have reported a frequency downshift.
This notable discrepancy is attributed to the difference in geometry and highlights peculiarities
of the two-dimensional problem which may not be widely known, as opposed to the three-dimensional case.

Possible future work includes a detailed phase plane analysis of solutions to the new RP equation,
exploiting its Hamiltonian structure and examining the role of the external pressure field.
In addition to mathematical results, the present study explores the possibility of using a boundary perturbation method
(as represented by the Taylor expansion of the DNO) to simulate the nonlinear surface dynamics of a gas bubble.
The focus here is on the algorithmic development and preliminary assessment of this numerical solver,
in cases where the processes of radial pulsations and shape distortions do not interact.
On the other hand, the nonlinear coupling between these different modes under resonant or near-resonant conditions
(e.g. under tuned harmonic forcing) constitutes an interesting problem in its own right, leading to complex phenomena \cite{gi18,ml99b}.
Its in-depth investigation via this numerical scheme is envisioned for a subsequent paper.

Finally, we emphasize that the proposed mathematical formulation and computational strategy 
are not restricted to the two-dimensional setting.
Provided a FFT version in spherical harmonics is available \cite{m99}, we plan to extend the present results
to the more general problem on nonlinear deformations of three-dimensional bubbles.
Related work in the context of three-dimensional water waves can be found in \cite{dp96,xg09}.

\section*{Acknowledgements}

P. Guyenne is partially supported by the NSF under grant No. DMS-2307712.

\clearpage

\clearpage

\end{document}